\documentclass[usenatbib,iop,apj,tighten,numberedappendix,twocolappendix]{aeb_emulateapj_2015}
\usepackage{amsmath}
\usepackage{amssymb}
\usepackage{xspace}
\usepackage[normalem]{ulem}
\usepackage{pifont}

\usepackage{natbib}

\usepackage[usenames,dvipsnames]{color}
\usepackage{hyperref}
\definecolor{darkblue}{rgb}{0.0,0.0,0.3}
\hypersetup{colorlinks,breaklinks,linkcolor=darkblue,urlcolor=darkblue,anchorcolor=darkblue,citecolor=darkblue}

\def\s{{\rm s}} 
\def\yr{{\rm yr}} 
\def\hr{{\rm hr}}

\def\m{{\rm m}} 
\def\mm{{\rm m}\m} 
\def\km{{\rm k}\m} 
\def\pc{{\rm pc}} 
\def\kpc{{\rm k}\pc} 

\def\erg{{\rm erg}} 

\def\Jy{{\rm Jy}} 

\def\muas{\mu{\rm as}} 

\newcommand\bmath[1] {\mbox{\boldmath{$\mathrm {#1}$}}}

\def\dk2{{d^2\!k}}

\def\grad{\bmath{\nabla}}

\def\AIC{{\rm AIC}}

\def\u{\bmath{u}}
\def\du{\bmath{\delta u}}

\begin{document}

\title{
  Modeling Seven Years of Event Horizon Telescope Observations\\
  with Radiatively Inefficient Accretion Flow Models
}

\newcommand{\perimeter}{1}
\newcommand{\waterloo}{2}

\newcommand{\hay}{3}
\newcommand{\cfa}{4}

\newcommand{\naoj}{5}
\newcommand{\tokyo}{6}
\newcommand{\cita}{7}
\newcommand{\mpifr}{8}

\author{
  Avery~E.~Broderick\altaffilmark{\perimeter,\waterloo},
  Vincent~L.~Fish\altaffilmark{\hay},
  Michael~D.~Johnson\altaffilmark{\cfa},
  Katherine~Rosenfeld\altaffilmark{\cfa},
  Carlos~Wang\altaffilmark{\waterloo},\\
  Sheperd~S.~Doeleman\altaffilmark{\hay,\cfa},
  Kazunori~Akiyama\altaffilmark{\naoj,\tokyo,\hay},
  Tim~Johannsen\altaffilmark{\cita,\waterloo,\perimeter},
  and Alan~L.~Roy\altaffilmark{\mpifr}
}
\altaffiltext{\perimeter}{Perimeter Institute for Theoretical Physics, 31 Caroline Street North, Waterloo, ON N2L 2Y5, Canada}
\altaffiltext{\waterloo}{Department of Physics and Astronomy, University of Waterloo, 200 University Avenue West, Waterloo, ON N2L 3G1, Canada}
\altaffiltext{\hay}{Massachusetts Institute of Technology, Haystack Observatory, Route 40, Westford, MA 01886, USA}
\altaffiltext{\cfa}{Harvard-Smithsonian Center for Astrophysics, 60 Garden Street, Cambridge, MA 02138, USA}
\altaffiltext{\naoj}{National Astronomy Observatory of Japan, Osawa 2-21-1, Mitaka, Tokyo 181-8588, Japan}
\altaffiltext{\tokyo}{Department of Astronomy, Graduate School of Science, The University of Tokyo, 7-3-1 Hongo, Bunkyo-ku, Tokyo 113-0033, Japan}
\altaffiltext{\cita}{Canadian Institute for Theoretical Astrophysics, University of Toronto, 60 St.\ George Street, Toronto, ON M5S 3H8, Canada}
\altaffiltext{\mpifr}{Max-Planck-Institut f\"{u}r Radioastronomie, Auf dem H\"{u}gel 69, D-53121 Bonn, Germany}

\shorttitle{Modeling Seven Years of EHT Data with RIAFs}
\shortauthors{Broderick et al.}

\begin{abstract}
  An initial three-station version of the Event Horizon Telescope, a millimeter-wavelength very-long baseline interferometer, has observed Sagittarius A* (Sgr~A*) repeatedly from 2007 to 2013, resulting in the measurement of a variety of interferometric quantities.  Of particular importance, there is now a large set of closure phases, measured over a number of independent observing epochs.  We analyze these observations within the context of a realization of semi-analytic radiatively inefficient disk models, implicated by the low luminosity of Sgr~A*.  We find a broad consistency among the various observing epochs and between different interferometric data types, with the latter providing significant support for this class of models of Sgr~A*.  The new data significantly tighten existing constraints on the spin magnitude and its orientation within this model context, finding a spin magnitude of $a=0.10^{+0.30+0.56}_{-0.10-0.10}$, an inclination with respect to the line of sight of $\theta={60^\circ}^{+5^\circ+10^\circ}_{-8^\circ-13^\circ}$, and a position angle of $\xi={156^\circ}^{+10^\circ+14^\circ}_{-17^\circ-27^\circ}$ east of north.  These are in good agreement with previous analyses.  Notably, the previous $180^\circ$ degeneracy in the position angle has now been conclusively broken by the inclusion of the closure phase measurements.  A reflection degeneracy in the inclination remains, permitting two localizations of the spin vector orientation, one of which is in agreement with the orbital angular momentum of the infrared gas cloud G2 and the clockwise disk of young stars.  This possibly supports a relationship between Sgr~A*'s accretion flow and these larger-scale features.
\end{abstract}


\section{Introduction}
The radio bright point source inhabiting the Galactic center, Sagittarius A* (Sgr~A*), is believed to be associated with an accreting supermassive black hole.  That Sgr~A* is a black hole is strongly supported by the dynamics of stars in its vicinity, which imply a central mass of $4.3\times10^6\,M_\odot$\footnote{Here we will assume a distance of $8~\kpc$.  The mass of Sgr~A* enters primarily via the determination of the angular scale subtended by the black hole, i.e., the combination $M/D$.  While the mass and distance estimates from stellar motions remain uncertain, these are strongly correlated such that $M/D$ is constrained to about 6\%.} lying completely within $0.01~\pc$ \citep{2008ApJ...689.1044G,2009ApJ...707L.114G}.  This is further evidenced by the lack of a significant galactocentric proper motion of the associated radio source \citep{2004ApJ...616..872R}.  That Sgr~A* is indeed a black hole, i.e., contains a horizon, is implied by its spectral energy distribution (SED), which lacks the thermal bump associated with accretion onto a photosphere \citep{2006ApJ...638L..21B, 2009ApJ...701.1357B}.

Millimeter-wavelength very-long baseline interferometry (mm-VLBI) provides the unique opportunity to directly probe spatial scales commensurate with the event horizon in Sgr~A*.  This is facilitated by two additional critical simplifications: Sgr~A*'s SED exhibits the characteristic transition from optically thick to optically thin near 1~mm, and the intervening electron scattering screen becomes sub-dominant below 1~mm.  As a result, at mm wavelengths VLBI is capable of resolving the near-horizon emission of Sgr~A*, probing its morphology and dynamics.

The Event Horizon Telescope (EHT) is a global mm-VLBI array comprised of existing mm wavelength facilities \citep{2009astro2010S..68D,2010evn..confE..53D}.  An early version of the EHT, consisting of stations in Hawaii (James Clerk Maxwell Telescope, JCMT; Submillimeter Array, SMA), Arizona (Arizona Radio Observatory Submillimeter Telescope, ARO-SMT), and California (Combined Array for Research in Millimeter-wave Astronomy, CARMA), detected sub-horizon scale structure in Sgr A* in 2007 \citep{2008Natur.455...78D}.  Since that time, the proto-EHT has continued to observe Sgr A* in 2011 and 2013 \citep{2011ApJ...727L..36F, FISH15}.  As a result, EHT observations extending over 16 nights spread over six years and consisting of both correlated flux densities and closure phases have been collected.

The sparseness of the baseline coverage, a result of the limited number of participating stations, has prevented the generation of an image directly from the interferometric data.  While such an ability is expected in the coming few years \citep{2014ApJ...795..134F}, it is unclear that this will be the optimal way in which to confront theoretical models of the accretion flow onto Sgr~A* for two reasons: first, the observational uncertainties are best understood in the interferometric data products specifically and second, the astrophysical modeling of Sgr~A* permits the inclusion of a wealth of additional information about the source.  Chief among the auxiliary data sets is the SED itself, followed by the implied limits on the emission region electron density resulting from the detection of mm-wavelength polarization \citep{2000ApJ...534L.173A, 2003ApJ...588..331B, 2006ApJ...640..308M, 2006JPhCS..54..354M, 2006PhDT........32M}.

Here we analyze the existing EHT data within the context of a class of radiatively inefficient accretion flow (RIAF) models that are consistent with the broad observational context of Sgr~A*.  These are characterized by hot, geometrically thick disks containing substantial disk-wind driven mass loss, motivated by the low electron density near the horizon and extended X-ray emission associated with Sgr A* \citep{2013Sci...341..981W}.  RIAFs comprise a large class of models, differing in choices for electron-ion coupling, transport properties, and assumed outflow efficiency, and include both semi-analytical and numerical solutions.  We use a specific choice based on the semi-analytical models described in \citet{2003ApJ...598..301Y}.  This improves on the analysis of \citet{2011ApJ...735..110B}, which was restricted to the correlated flux densities in 2007 and 2009, by including closure phase measurements made from 2009 through 2013.

The new data set enables us to address two issues: First, the results of \citet{2011ApJ...735..110B} are a strong prior on the kinds of RIAF models that are applicable to Sgr~A*, and therefore the subsequent data provides a critical test of the RIAF picture.  More importantly, the additional data is in the form of closure phases, which are distinct from and considerably more sensitive to the emission region structure than the correlated flux densities considered in \citet{2011ApJ...735..110B}, and hence there is no a priori expectation that the preferred models from \citet{2011ApJ...735..110B} will fit the new closure phase data at all.  Second, the substantial increase in total observation time permits a corresponding improvement in the parameter estimation, i.e., the reconstruction of the black hole spin magnitude and orientation.

We summarize the EHT data employed in Section \ref{sec:data}.  The RIAF modeling of the interferometric data is presented in Sections \ref{sec:models} and \ref{sec:CIO}.  The consistency with \citet{2011ApJ...735..110B} is assessed and updated parameter estimates are presented in Section \ref{sec:results}.  Finally, discussion and conclusions are contained in \ref{sec:conclusions}.

\begin{deluxetable}{ccccccc}
\tablecaption{Data Epochs\label{tab:data}}
\tablehead{Epoch & Year & Day(s) & Time & N\tablenotemark{a} & Type\tablenotemark{b} & Ref\tablenotemark{c}}
\startdata
1      & 2007 & 100-101 & 11.00-13.67 & 19 & VM & D8\\
2      & 2009 & 95      & 11.17-15.00 & 12 & VM & F11\\
3      & 2009 & 96      & 11.50-14.56 & 19 & VM & F11\\
4      & 2009 & 97      & 11.50-13.67 & 20 & VM & F11\\
Totals & ...  & ...     & 11.73 hrs   & 70 &    &\\
\hline\\
5      & 2009 & 93      & 11.54-13.87 & 11 & CP & F15\\
6      & 2009 & 96      & 12.46-12.79 & 3  & CP & F15\\
7      & 2009 & 97      & 11.96-14.38 & 10 & CP & F15\\
8      & 2011 & 88      & 12.37-13.52 & 7  & CP & F15\\
9      & 2011 & 90      & 13.67-14.02 & 2  & CP & F15\\
10     & 2011 & 91      & 11.93-13.53 & 5  & CP & F15\\
11     & 2011 & 94 & 11.78-14.51 & 17 & CP & F15\\
12     & 2012 & 81      & 12.52-15.68 & 25 & CP & F15\\
13     & 2013 & 80      & 12.55-15.43 & 28 & CP & F15\\
14     & 2013 & 81      & 12.97-15.27 & 10 & CP & F15\\
15     & 2013 & 82      & 12.97-14.88 & 15 & CP & F15\\
16     & 2013 & 85      & 12.15-15.17 & 32 & CP & F15\\
17     & 2013 & 86      & 12.55-13.95 & 16 & CP & F15\\
Totals & ...  & ...     & 25.58 hrs   & 181 &   & \\
\enddata
\tablenotetext{a}{Number of data points, including detections only}
\tablenotetext{b}{Data types are visibility magnitudes (VM) and closure phases (CP)}
\tablenotetext{c}{D8=\citet{2008Natur.455...78D}, F11=\citet{2011ApJ...727L..36F}, F15=\citet{FISH15}}
\end{deluxetable}

\section{EHT Data} \label{sec:data}

\begin{figure}
\begin{center}
\includegraphics[width=\columnwidth]{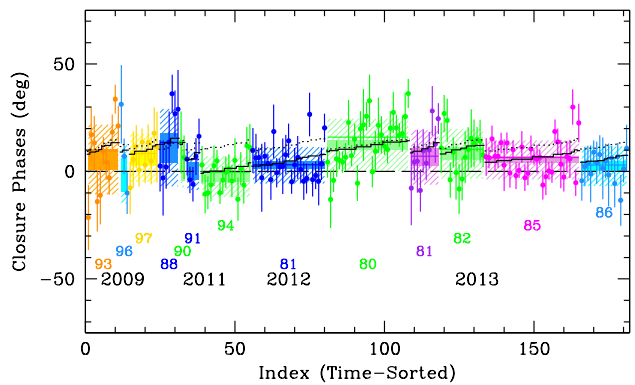}
\end{center}
\caption{Each of the 181 closure phases measurements for Sgr A* in time order with the errors assumed here.  For reference the hatched and filled regions show the day-specific estimates of the scatter and medians from the smoothed bootstrap analysis.  The theoretically anticipated closure phases arising from our most probable RIAF model are shown with (solid) and without (dotted) the inclusion of $\bar{\phi}_E$.  Note that the step-like nature of this line is due to degenerate baseline triangles in the data.  This figure may be directly compared to Figure 2 of \citet{FISH15}.} \label{fig:CPs}
\end{figure}

We employ the results of a large number of 1.3~mm-VLBI experiments conducted from 2007 through 2013 using stations in Hawaii (JCMT, SMA, CSO) and the continental United States (CARMA, SMT) under the auspices of the EHT project.  The corresponding data set consists of two distinct interferometric products, probing the amplitude and phase structure of the complex visibilities.  Individual measurements correspond to scans, typically of 5--15 minute length, with a typical cadence of approximately 20--30 minutes over the 4--5 hours that Sgr~A* is mutually visible at all US stations.  We collect these into epochs, one for each night of observations (with the exception of the first, which is comprised of two nights), allowing us to assess the variability and inter-epoch consistency of the model fits.  Note that the closure phases reported on day 94 of 2011 (epoch 11) appear anomalously low; this statement is strongly dependent on the specifics of the fringe search method applied \citep{FISH15}.  The full data set is comprised of 17 data epochs, listed in Table \ref{tab:data}, and represent a contiguous observing time of 35.27~hr.

\subsection{Visibility Magnitudes}

The amplitudes of the complex visibilities are weakly dependent on the potentially large phase delays that occur during propagation through the atmosphere.  While they are impacted by atmospheric absorption and gain uncertainties within the telescopes, this can be addressed via careful calibration.  Encoded in the amplitudes is primarily information on characteristic scales within the emission region.  More importantly, these can be constructed from observations by pairs of telescopes, permitting probes of horizon-scale structures without multiple long baselines.  Hence, the initial mm-VLBI observations reported visibility magnitudes exclusively. Here we employ the observations reported in \citet{2008Natur.455...78D} and \citet{2011ApJ...727L..36F}, in which the full details of the observations, calibration, and data processing can be found.

\citet{2008Natur.455...78D} describes mm-VLBI observations conducted over two nights in April, 2007, using the JCMT, SMT, and a single CARMA antenna.  Nineteen visibility amplitudes were measured on the CARMA-SMT and JCMT-SMT baselines, with only an upper limit obtained on the JCMT-CARMA baseline, which we ignore in favor of more recent detections due to the weak constraint it applies.  Signal-to-noise ratios of the incoherently averaged visibility magnitudes of 4 and 8 were typical on the short and long baselines, respectively.  The full CARMA array was operated concurrently as a stand-alone array, and measured an effective zero-baseline flux of $2.4\pm0.25~\Jy$.  This is similar to the visibility magnitudes measured on the shorter CARMA-SMT baseline.  It is, however, anomalously low in comparison to the more typical 3~Jy flux at 1.3~mm, implying that Sgr~A* was in a quiescent state.

\citet{2011ApJ...727L..36F} present subsequent observations over 3 nights in April, 2009 that employed the JCMT, SMT, and multiple CARMA antennas operated independently.  A total of fifty-four visibility magnitudes were obtained on JCMT-SMT, CARMA-SMT, and both JCMT-CARMA baselines on two of the three days.  Signal-to-noise ratios of the incoherently averaged visibility amplitudes were considerably higher, reaching 17 and 5 on the short and long baselines, respectively.  In addition the two independent CARMA antennae formed a very short baseline, accessing angular scales of order 10'' and finding substantially more correlated flux than implied by detections on the CARMA-SMT baselines.  As in \citet{2011ApJ...735..110B}, we will assume that this arises from a separate large-scale component not present in the 2007 observations.  This is supported by the similarities in the structures inferred using only the 2007 and 2009 visibility magnitudes despite significant variations in their over-all normalizations \citep{2011ApJ...727L..36F}.  Hence, we do not consider the inter-CARMA visibilities further here.

Because of the challenges in calibrating visibility magnitudes and the novel nature of closure phases (see the following section) we leave the consideration of visibility magnitudes obtained in subsequent observing runs for future work.

\begin{figure}
  \begin{center}
    \includegraphics[width=\columnwidth]{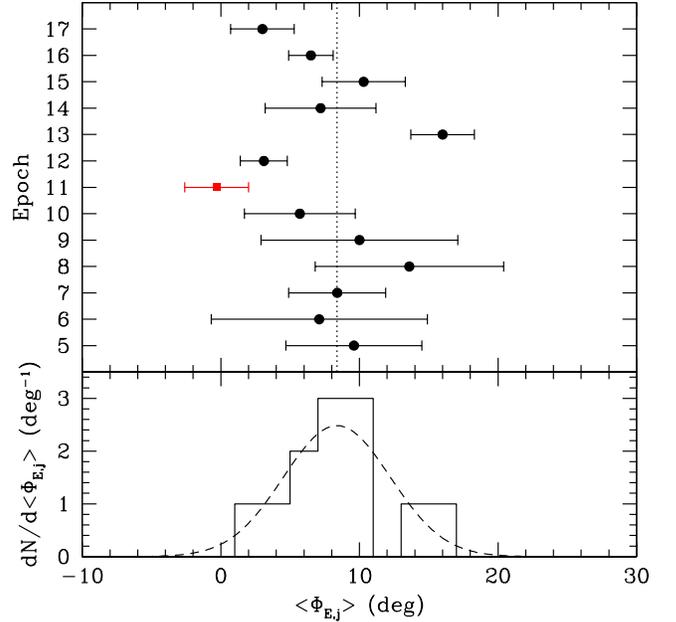}
  \end{center}
  \caption{Top: Closure phase means for each epochs considered here, including the uncertainty estimated from bootstrap sampling \citep{FISH15}.  The red square corresponds to epoch 11, for which the reconstructed closure phase is anomalously low and sensitive to the fringe search method employed.  For this reason it has been excluded from the estimation of the intraday closure phase variations.  Bottom: The resulting distribution is reasonably well fit by a normal distribution with mean $8.4^\circ$ (dotted line in the top panel) and standard deviation $3.86^\circ$, shown by the dashed curve. After subtracting the mean, we adopt this distribution as the prior on the potential closure phase offsets, $\phi_E$.}
\label{fig:CPdist}
\end{figure}

\subsection{Closure Phases}

The measured phase of a complex visibility, $\Phi_E$, contains information on the structure of the observed source but is corrupted by variations introduced by propagation of radiation through the atmosphere.  At longer wavelengths these delays can be removed by rapid nodding between a target source and a nearby calibrator, but at 1.3~mm the timescale over which the atmospheric phase contribution changes by a radian can be too rapid---as short as a few seconds depending on weather conditions---to permit phase transfer from a nearby calibrator.

Fortunately, it is possible to construct VLBI observables that depend on the visibility phases of a source while being robust against atmospheric corruption.  The simplest such quantity, closure phase, is constructed by taking the directed sum of the visibility phases along a closed triangle of baselines.  The phase introduced by the unknown atmospheric delay at a station on one baseline is exactly cancelled by the phase on the other baseline including that station.  In fact, closure phases are insensitive to almost all station-based phase effects, whether atmospheric or instrumental in origin.

The closure-phase data set used in this work consists of 181 measurements on the California-Hawaii-Arizona triangle taken from 2009 March through 2013 March.  The data are described in greater detail in \citet{FISH15}.  These are shown explicitly in Figure \ref{fig:CPs} as a function of index, which indicates (roughly) the time ordering.  As with the amplitude data, errors on the closure phase data are smaller in later epochs due to increases in sensitivity of the array, especially the use of phased arrays at the SMA and CARMA.  An analysis of the closure phase data determined that the error distributions are very well approximated by Gaussians with a standard deviation in radians equal to the inverse of the signal-to-noise ratio.

\section{RIAF Modeling} \label{sec:models}
We employ the same library of radiatively inefficient accretion flow models presented in \citet{2011ApJ...735..110B}, to which we direct the reader for more detail.  These are based on the one-dimensional models of \citet{2003ApJ...598..301Y}, and are characterized by geometrically thick, nearly virialized ion distributions with a substantially sub-equipartition electron population, a result of the weak coupling between the two species at the low accretion flow densities appropriate for Sgr~A*.  The radial spatial distributions of the densities and temperatures of the plasma components is assumed to be power laws, similar to the semi-analytical results of \citet{2003ApJ...598..301Y}.  In particular, substantial wind loss is assumed to produce the small near-horizon densities implied by spectropolarimetry \citep{2006ApJ...640..308M,2003ApJ...588..331B, 2000ApJ...538L.121A}.

We assume the accretion flow is orbiting at the Keplerian velocity beyond the innermost stable circular orbit (ISCO), and plunging on constant angular momentum orbits inside of the ISCO.  In all cases we assume that the angular momentum of the accretion flow is aligned with the black hole spin.  Finally, we assume a dominantly toroidal magnetic field with a normalization set at 10\% of equipartition with the ions.

The emission is due to synchrotron radiation, associated with a multi-component population of relativistic electrons, consisting of a ``thermal'' and power law energy distributions, with the latter cutting off below Lorentz factors of $10^2$.  The index of the power law component is set by the optically-thin millimeter to near-infrared SED of Sgr~A*, while its radial structure is determined by the self-absorbed radio SED.  We perform the full polarized self-absorbed radiative transfer along null geodesics in the covariant formulation described in \citet{2004MNRAS.349..994B, 2006MNRAS.366L..10B}.

The normalizations of the electron (and therefore ion) distributions are set by fitting the observed SED of Sgr~A*.  These are necessarily functions of the viewing geometries and black hole parameters, and thus a library of solutions are obtained as a function of dimensionless spin magnitude\footnote{The black hole angular momentum is related to the dimensionless spin magnitude via $J=a GM^2/c$.  Note that often $J/Mc$ is called the ``spin'', though here we will use this interchangeably with $a$.}, $a$, and viewing inclination, $\theta$.  Modifications to the position angle, $\xi$, measured east of north, are affected by rotations of the resulting images in the plane of the sky.

Note that these are models of the quiescent accretion flow, and thus ignore variability.  This includes the expected small-scale variations due to turbulence, currently believed necessary to drive angular momentum transport.  We will, however, attempt to partially include the impact of small-scale brightness fluctuations in the image in a systematic correction discussed in the following section.

The model library is then interpolated to produce a 1.3mm image library containing 9090 images spanning spins from 0 to 0.998 and inclinations from $0^\circ$ to $90^\circ$.  Rotations and reflections are used to cover the remaining inclination parameter space.

\section{Computing Interferometric Observables}\label{sec:CIO}
Visibilities are computed from the theoretically produced images as described in \citet{2011ApJ...735..110B}.  This is complicated by the presence of an interstellar electron scattering screen in the direction of the Galactic center which effectively blurs the image.  In practice, the visibility computation is performed in two steps.  First the complex visibilities are constructed from the unscattered theoretical images in the normal way.  Second we perform the scattering convolution in the Fourier domain.  The two interferometric observables, visibility magnitudes and closure phases, are then constructed from the theoretical complex visibilities as described in \citet{2011ApJ...735..110B, 2011ApJ...738...38B} for the relevant locations and triangles in the $u$-$v$ plane.  Note that within the ensemble-average regime, wherein the scattering can be treated as a Gaussian convolution, it does not have any impact on the closure phases.

The scattering is both wavelength dependent and anisotropic, becoming subdominant at mm wavelengths, though not negligible.  Nevertheless, it is well approximated on the scales of interest by a convolution with an asymmetric Gaussian kernel, whose orientation, size, and wavelength have been empirically determined at long wavelengths \citep{2006ApJ...648L.127B}.  The convolution approximation is only strictly appropriate when images are averaged over a long period of time (many days), and the effects of scattering on single-epoch images can introduce small inter-epoch shifts in the closure phases.  We attempt to partially account for these below.

For analysis the data is collected into 17 epochs, with epoch 1 consisting of all the 2007 data due to its lower precision and all remaining epochs consisting of data from a single observation night, consisting of roughly 2~hr observing runs (see Table \ref{tab:data}).  Sgr~A* exhibits both sporadic short-term variability, e.g., flare activity,  and long-term variability, e.g., changes in accretion rate.

On day timescales the flux variations do not appear to be associated with large-scale structure changes in the source \citep{2011ApJ...727L..36F}, and thus we model these via small changes in the instantaneous mass accretion rate.  As a result, we have allowed for a linear flux renormalization of the theoretical images to vary independently for epochs 1--4, impacting the visibility magnitudes; closure phases are unchanged by a linear renormalization of the flux.

However, closure phases are impacted by inter-epoch variability in the small-scale image structure \citep[e.g.,][]{2009ApJ...695...59D}.  Here we assume that the inter-epoch variations in the closure phases are the result of small-scale brightness fluctuations due either to refraction in the intervening scattering screen \citep[e.g.,][]{0004-637X-805-2-180} or turbulence within the accretion flow itself.  In both cases the result during quiescent periods is to induce small shifts in the closure phases that are stable over hours to days (see Appendix \ref{app:CPvar} and Figure \ref{fig:CPdist}).\footnote{Note that we necessarily cannot model short-timescale variability, e.g., flaring emission, in this way.  The sub-hour scale image variability will produce fluctuations in the associated visibilities on comparable timescales \citep[e.g.,][]{2009ApJ...695...59D,2009ApJ...706.1353F} and require sub-epoch modeling.}  Thus, we introduce an additional twelve parameters, $\phi_E$, corresponding to the epoch-specific closure phase shifts.  (This is similar to the epoch-specific flux renormalization employed for epochs 1-4.)  To prevent large shifts that are inconsistent with the assumed physical mechanisms underlying the closure phase variations we adopt a Gaussian prior on $\phi_E$ with a width equal to the observed mean closure phase distribution after excluding epoch 11, for which the apparently anomalous nature of its low value is highly dependent on the details of the data analysis \citep{FISH15}, i.e., $\sigma_\phi=3.86^\circ$ (see Figure \ref{fig:CPdist}).  Note that apart from the prior, permitting shifts in the closure phase data substantially weakens the probative value of the mean closure phases -- these are now assumed to be contaminated by the poorly constrained small-scale structures in the image.  However, the structure constraint inherent in the evolution of the closure phases remains.

While the processes we adopt to explain the closure phase variations are astrophysically reasonable, it remains unclear that these are the cause in practice.  Therefore, we also analyze the entire EHT data set setting the $\phi_E=0$, in which case the systematic shifts in the closure phases are treated as random errors.  While agnostic regarding the source of the variation, it is clear at the outset that a single quiescent image structure is formally inconsistent with the set of closure phases measured.  This conclusion is validated in the subsequent analysis -- some additional systematic must be invoked to obtain satisfactory fits.  However, as we discuss in the following section, such a model is strongly disfavored.

\begin{deluxetable}{cccccccc}
\tablecaption{Fit Results by Epoch\label{tab:chi2e}}
\tablehead{
  Epoch &
  N\tablenotemark{a} &
  k\tablenotemark{b} &
  min $\chi^2$\tablenotemark{c} &
  fit $\chi^2$\tablenotemark{d} &
  $V_{00}$\tablenotemark{e} &
  $\phi_E^M$\tablenotemark{f} &
  $\bar{\phi}_E$\tablenotemark{g}
}
\startdata
       1 &       19 &        4 &     5.77 &      7.5 &     2.44 &      ... &      ... \\ 
       2 &       12 &        4 &     10.8 &       11 &     2.16 &      ... &      ... \\ 
       3 &       19 &        4 &     18.8 &     28.3 &     2.12 &      ... &      ... \\ 
       4 &       20 &        4 &     12.4 &     13.4 &     2.95 &      ... &      ... \\ 
       5 &       11 &        4 &     9.68 &     22.3 &      ... &     1.90 &     1.21 \\ 
       6 &        3 &        4 &     1.63 &     4.19 &      ... &     9.10 &     2.51 \\ 
       7 &       10 &        4 &      2.9 &     3.96 &      ... &     4.02 &     2.42 \\ 
       8 &        7 &        4 &    0.883 &     8.41 &      ... &    -5.92 &    -2.37 \\ 
       9 &        2 &        4 & 1.74e-12 &    0.185 &      ... &     5.05 &    0.831 \\ 
      10 &        5 &        4 &     1.45 &     3.73 &      ... &     8.34 &     4.53 \\ 
      11 &       17 &        4 &     10.0 &     11.4 &      ... &     13.6 &     10.9 \\ 
      12 &       25 &        4 &     14.5 &     23.1 &      ... &     8.30 &     7.24 \\ 
      13 &       28 &        4 &     27.5 &     44.7 &      ... &     0.73 &     0.65 \\ 
      14 &       10 &        4 &     3.77 &     9.05 &      ... &     2.91 &     1.76 \\ 
      15 &       15 &        4 &     8.62 &     14.2 &      ... &     3.54 &     2.55 \\ 
      16 &       32 &        4 &     37.8 &     46.9 &      ... &     5.61 &     5.23 \\ 
      17 &       16 &        4 &     4.83 &        6 &      ... &     7.91 &     5.99
\enddata
\tablenotetext{a}{Number of data points, including detections only.}
\tablenotetext{b}{Number of fit parameters.}
\tablenotetext{c}{Epoch specific minimum $\chi2$.}
\tablenotetext{d}{Epoch specific $\chi2$ of the most probable model.}
\tablenotetext{e}{Visibility magnitude normalization.}
\tablenotetext{f}{Most likely closure phase offset.}
\tablenotetext{g}{Marginalized closure phase offset.}
\end{deluxetable}

\begin{deluxetable}{ccccccccc}
\tablecaption{Model Comparison\label{tab:chi2}}
\tablehead{
  Model Class &
  $N$\tablenotemark{a} &
  $k$\tablenotemark{b} &
  min $\chi^2$ &
  $\Delta\AIC$\tablenotemark{c} &
  Odds Ratio\tablenotemark{c}
}
\startdata
$\theta\ge90^\circ$ with $\phi_E=0$   & 251 & 7  & 336.4 & 57.0 & $9.7\times10^{-9}$ \\
$\theta\le90^\circ$ with $\phi_E=0$   & 251 & 7  & 338.8 & 59.5 & $6.2\times10^{-10}$ \\
$\theta\ge90^\circ$ with $\phi_E\ne0$ & 251 & 20 & 251.1 & 0.94 & 0.93   \\
$\theta\le90^\circ$ with $\phi_E\ne0$ & 251 & 20 & 250.2 & 0    & 1 \\
\enddata
\tablenotetext{a}{Number of data points, including detections only}
\tablenotetext{b}{Number of fit parameters}
\tablenotetext{c}{The $\theta\le90^\circ$, $\phi_E\ne0$ is used as the reference.}
\end{deluxetable}

\begin{figure*}
\begin{center}
\includegraphics[width=0.49\textwidth]{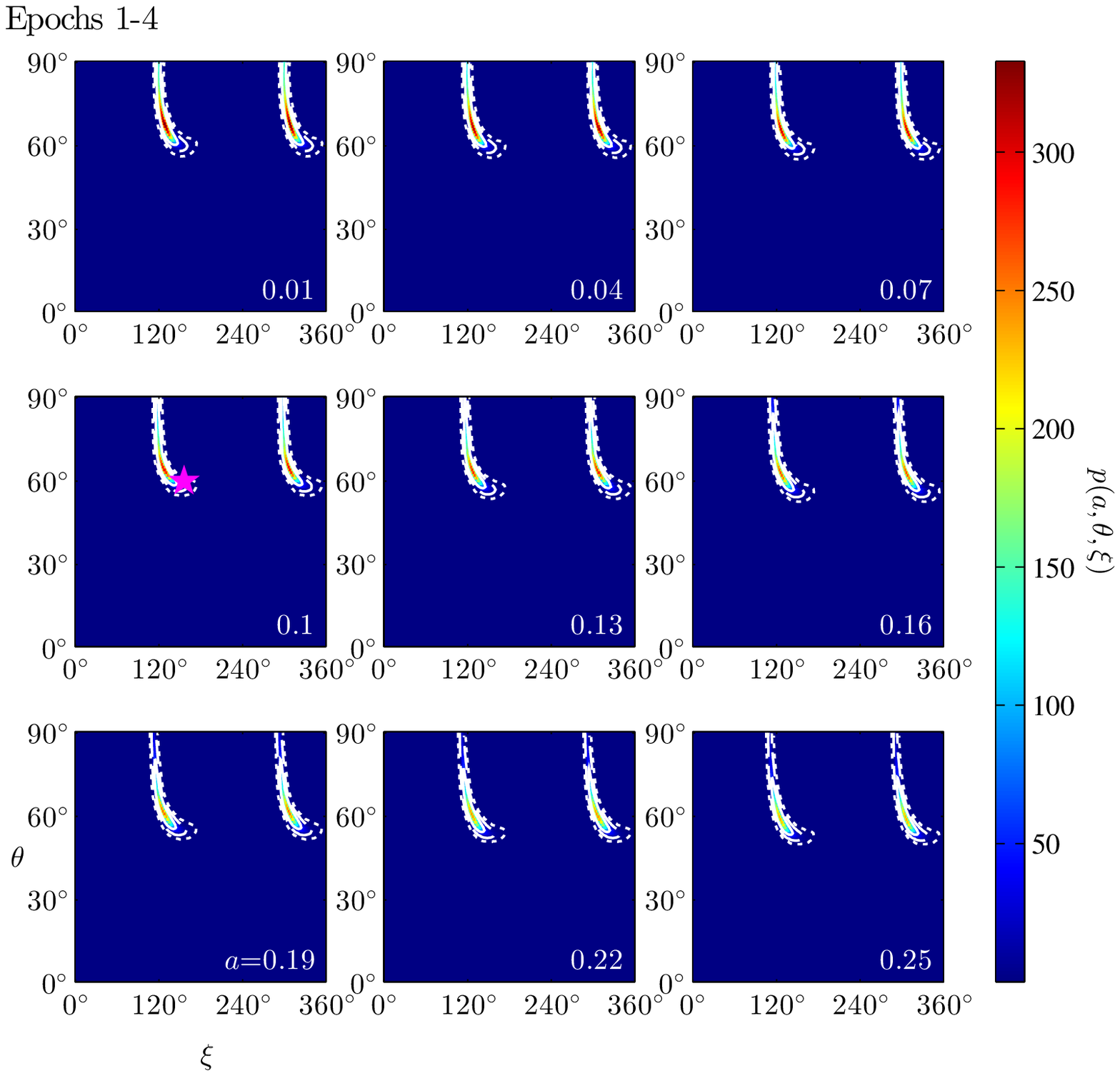}
\includegraphics[width=0.49\textwidth]{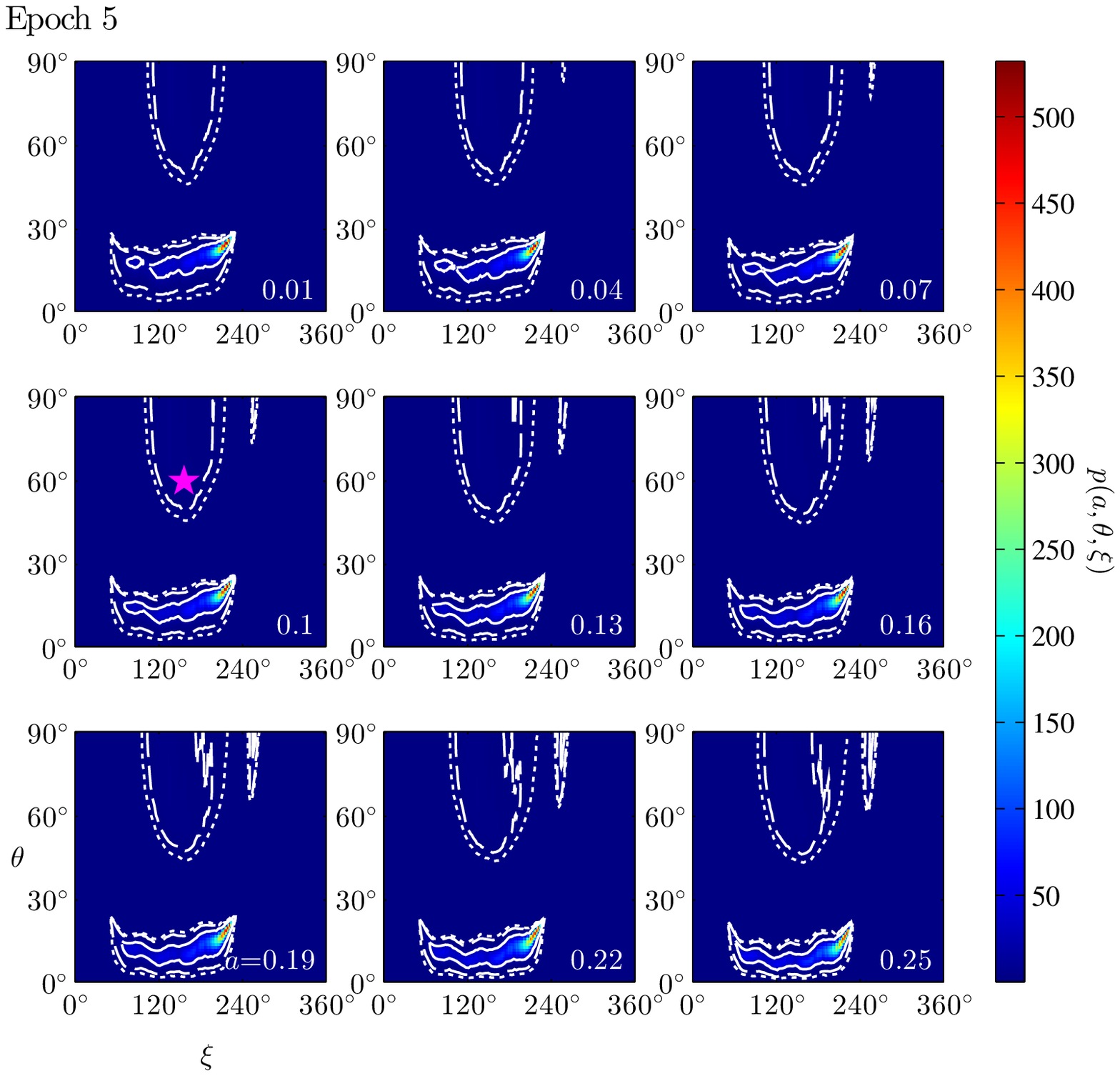}\\
\includegraphics[width=0.49\textwidth]{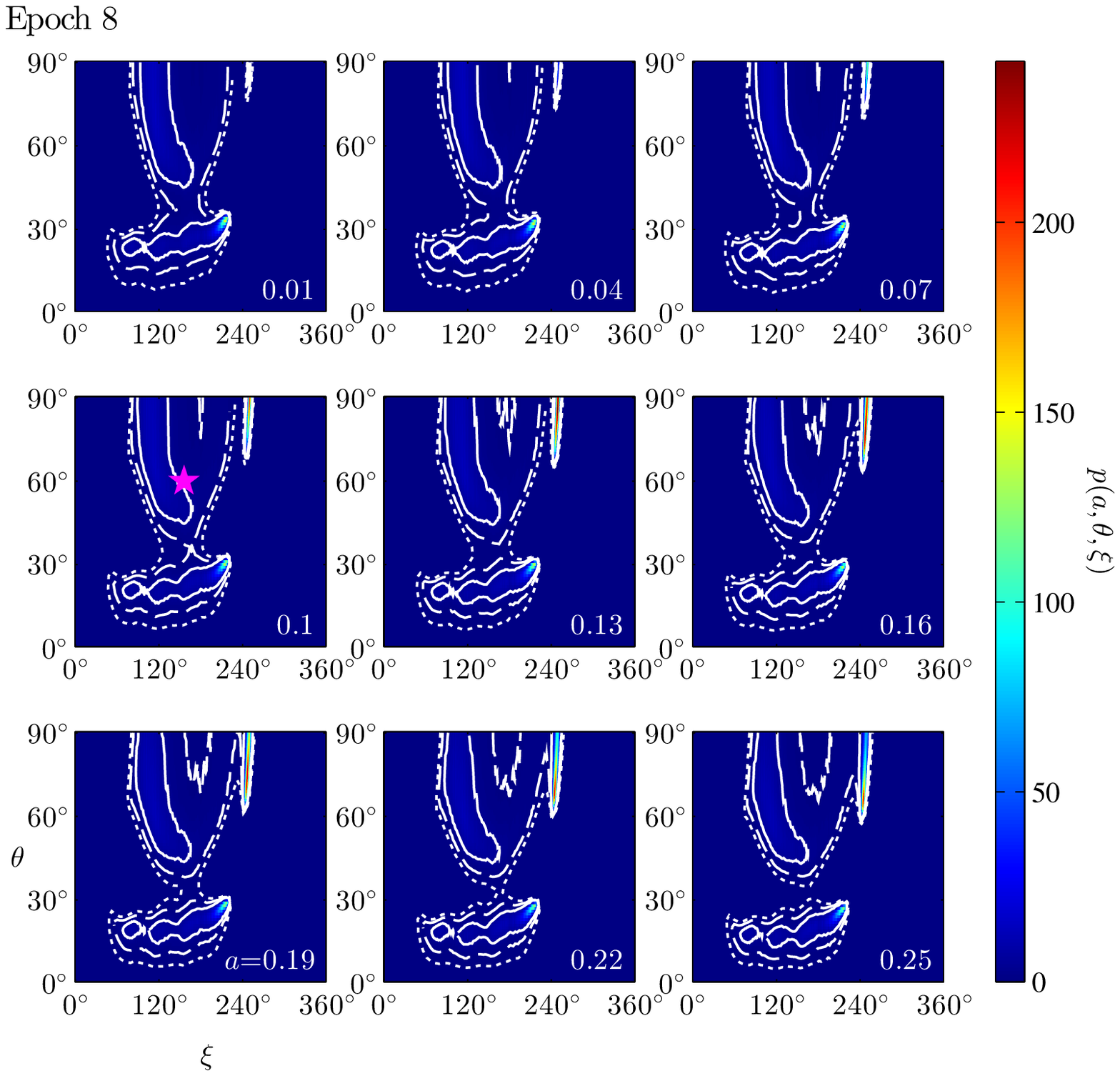}
\includegraphics[width=0.49\textwidth]{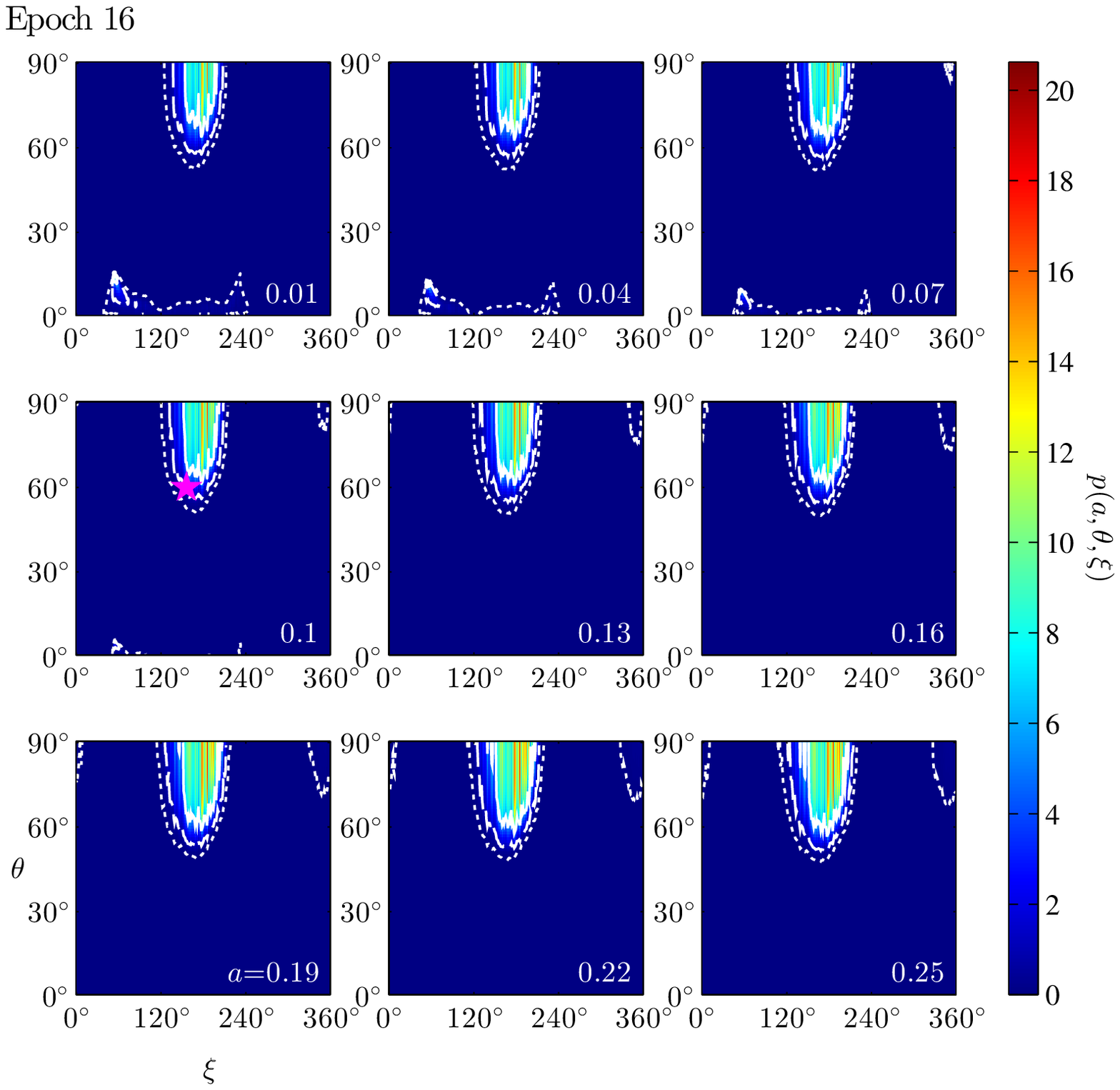}
\end{center}
\caption{Posterior probability distribution of the vector black hole spin.  Each panel shows a slice of constant $a$, listed in the lower left corner.  The solid, dashed, and dotted contours show the 1$\sigma$, 2$\sigma$, and 3$\sigma$ confidence regions, respectively.  The epoch number is listed in the bottom right panel, followed by the corresponding number of data points.  The best fit solution to all epochs is shown by the magenta star, located in the middle-left panel of each epoch-specific set.}\label{fig:Pm3Depochs}
\end{figure*}

\section{Results}\label{sec:results}

We seek to address three distinct statistical questions:
\begin{enumerate}
\item Do good fits exist?
\item Are the fits consistent across epochs?
\item What are the best fit parameters?
\end{enumerate}
For all three points the first step is the construction of observing epoch-specific likelihoods.  In doing this we assume Gaussian uncertainties for both the visibility magnitudes and closure phases (that this is well justified is shown in \citealt{FISH15}).  What happens afterward depends on the particular question being addressed; we consider each in turn here.

\subsection{RIAF Fit Quality}
We begin by assessing if our RIAF model class provides a statistically adequate description of the EHT observations.  We do this via the $\chi^2$ statistic, both for the entire data set and for contributions from sub-regions within it, listed in Tables \ref{tab:chi2e} and \ref{tab:chi2}.

For the entire data set the minimum $\chi^2$ is related to the maximum likelihood in the standard way and has the advantage of being easily interpreted.  We find a value of $250.2$, consistent with the expected value of $231$ at the 2$\sigma$ level, i.e., at a p-value of $18.4\%$.\footnote{The p-value is the probability of finding a $\chi^2$ in excess of the value obtained, assuming that the underlying model is correct.  This provides a quantitative measure of the significance of a given $\chi^2$ fluctuation, and therefore the need for additional model components: small p-values imply that the $\chi^2$ obtained is inconsistent with the expected statistical fluctuations and thus modifications to the model are required.}  Thus, as found in earlier analyses, RIAF models continue to provide an excellent fit to the horizon-scale structure of Sgr~A*.  Critically, note that the phenomenological models considered in \citet{2011ApJ...735..110B} as well as annuli and symmetric double point sources are unable to produce the small but non-zero closure phase evolutions observed, excluding these at overwhelming confidence\footnote{Note that it remains possible for more complicated multicomponent models to provide an adequate fit to the clsoure phase data, as described in detail in \citet{FISH15}.}.

A similar analysis can be performed for each epoch individually, assessing if the RIAF model provides a satisfactory fit for each.  This is complicated by the large variation in the number of data points within each epoch; some epochs have too few measured values to permit an independent fit, i.e., $N_e\le4$.  Excluding these from consideration, the remaining epochs all admit good fits, with the smallest p-value being $10\%$, occurring for epoch 16.  We also show the epoch-specific $\chi^2$ values after including an isotropic prior on the orientation and priors on $\phi_E$.  While marginally worse than the best fit models at each epoch, it is also a good fit for all epochs individually (minimum p-value of $2.2\%$, assuming the number of degrees of freedom is simply $N_e$).  Therefore, we conclude that there is no evidence that the RIAF model is insufficient to describe the quiescent emission of Sgr~A* for any epoch individually or for all epochs in combination.

\begin{figure*}
\begin{center}
\includegraphics[width=0.32\textwidth]{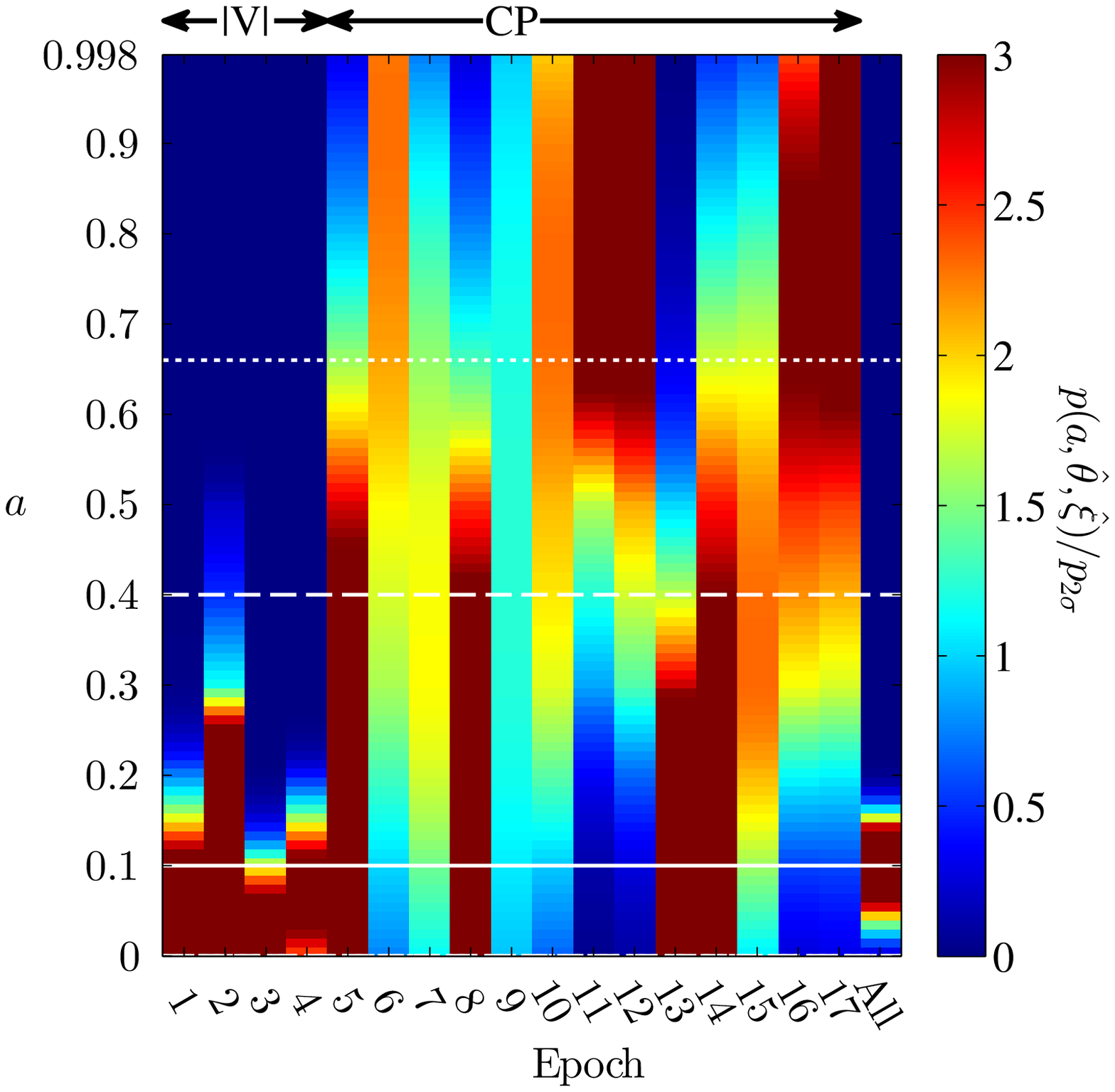}
\includegraphics[width=0.32\textwidth]{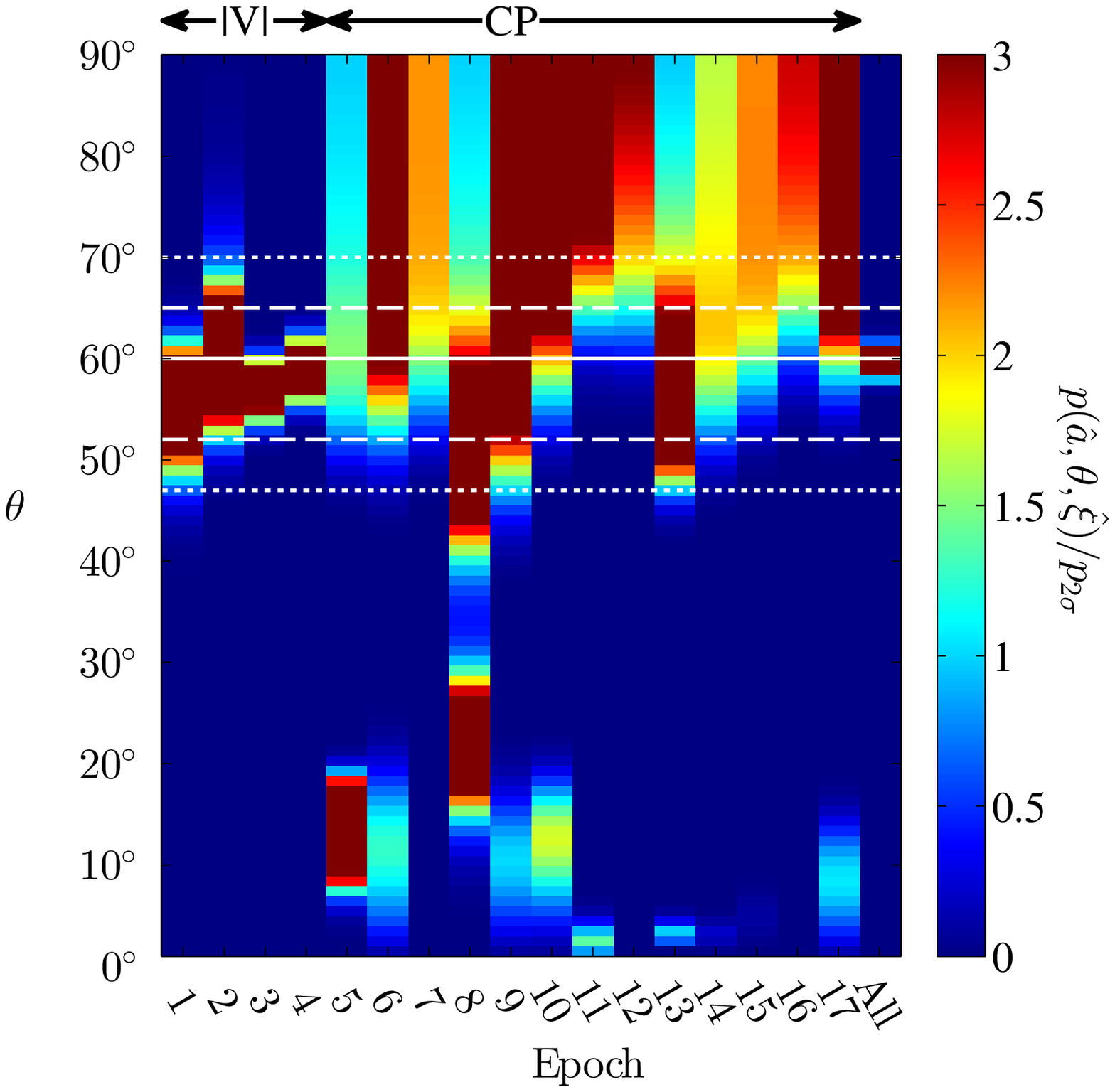}
\includegraphics[width=0.32\textwidth]{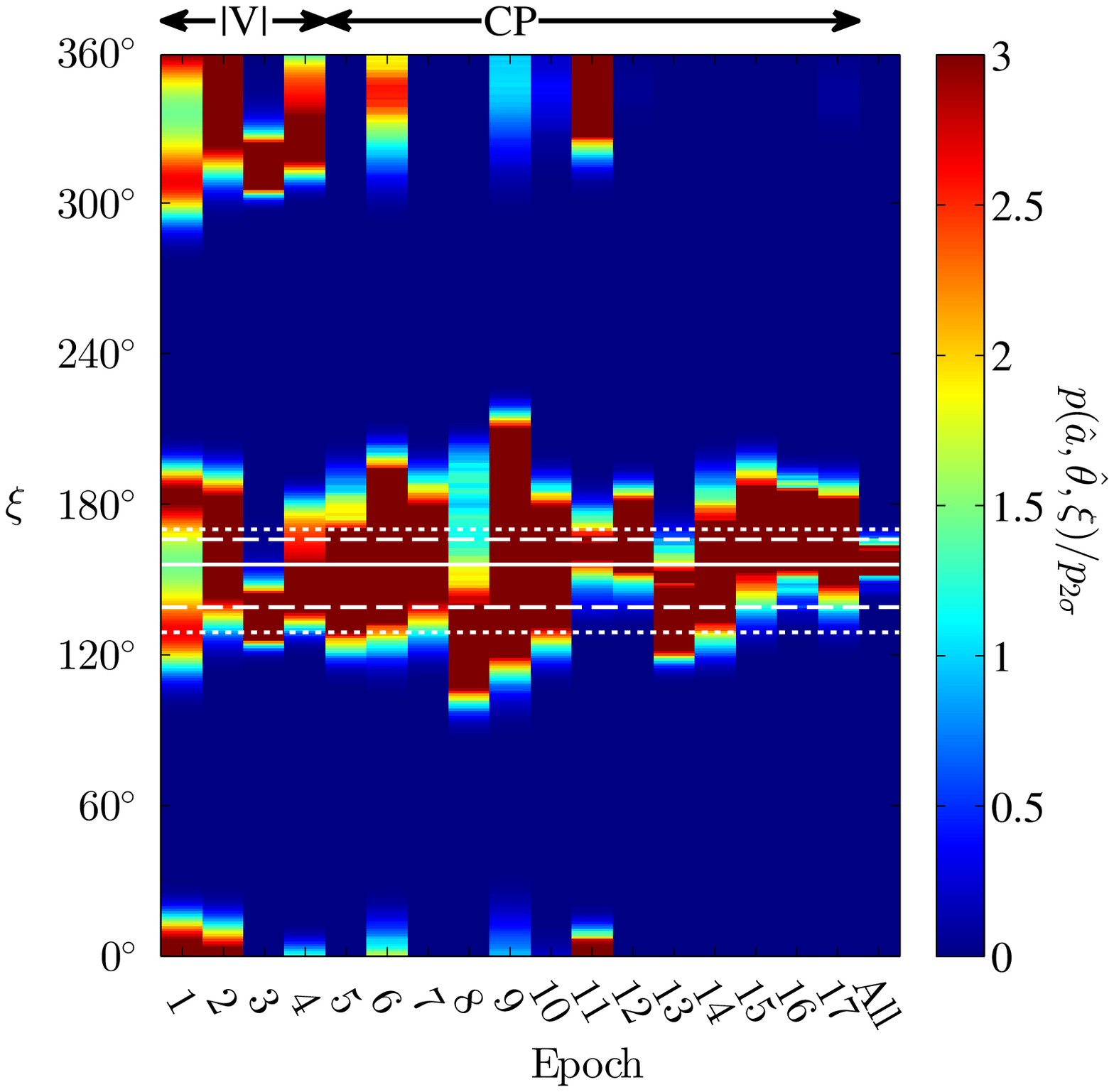}
\end{center}
\caption{Posterior probability as a function of epoch and spin magnitude (left), inclination (center), and position angle (right), along a chord through the three-dimensional spin parameter space at the best fit values of the remaining parameters.  The 1$\sigma$, 2$\sigma$, and 3$\sigma$ regions for each epoch, as defined by the cumulative probability, are shown by the colors.  The 1$\sigma$, 2$\sigma$, and 3$\sigma$ regions in the full spin parameter space, shown explicitly in Figure \ref{fig:Pm3Datx}, are given by the solid, dashed, and dotted lines, respectively.}\label{fig:Pm1Depoch}
\end{figure*}

\subsection{Closure Phase Fluctuations} \label{sec:CPFs}
Included in Table \ref{tab:chi2} are fit results for when the inter-epoch closure phase variations are not modeled.  As anticipated the quality of the resulting fits is much worse -- the minimum $\chi^2$ is $336.4$, with an associated p-value of $4.8\times10^{-6}$ or excluded at a level of more than $4.4\sigma$.  However, the smaller set of fit parameters, 7 instead of 20, motivates a careful comparison of the fit quality.  We do this in two ways, both of which are shown in Table \ref{tab:chi2}.  In all cases we take the $\theta<90^\circ$, with closure phase fluctuations modeled as our default case, to which the others are compared.

The first employs the Akaike information criterion ($\AIC$), defined by
\begin{equation}
\AIC = \chi^2_{\rm min} + 2k + \frac{2k(k+1)}{N-k-1}\,.
\end{equation}
where $k$ is the number of fit parameters, $N$ is the total number of data points, and $\chi^2_{\rm min}$ is the minimum $\chi^2$ achieved \citep{1974ITAC...19..716A,2000Ap&SS.271..213T,2002Burnham,2007MNRAS.377L..74L}.  The $\AIC$ is a penalized $\chi^2$ statistic and presents an approximate measure of the difference between the true and modeled data distributions \citep{2000Ap&SS.271..213T}.  Only differences in the $\AIC$ are statistically relevant, measured in the Jeffrey's scale, for which $\Delta\AIC>10$ are ``decisive'' evidence against the model with the higher $\AIC$.  The model which ignores the closure phase fluctuations is therefore decisively excluded by this criterion, exhibiting $\Delta\AIC>57$ for both potential inclinations relative to the models in which the inter-epoch closure phase variations are modeled.

The second measure we employ is the Bayesian Odds Ratio, defined by ratio of the model likelihoods, marginalized over all fit parameters \citep[see, e.g.,][]{2005blda.book.....G}.  That is given two models, $M_{I}$ and $M_{II}$, with parameters $\bmath{p}_{I,II}$, parameter priors $\Pi_{I,II}(\bmath{p}_{I,II})$, and associated likelihoods $L_{I,II}(\bmath{p}_{I,II})$, the Odds Ratio is given by
\begin{equation}
{\rm Odds~Ratio}
\equiv 
\frac{
  \int d^m\!p_I \,L_I(\bmath{p}_I) \,\Pi_I(\bmath{p}_I)
}{
  \int d^n\!p_{II} \,L_{II}(\bmath{p}_{II}) \,\Pi_{II}(\bmath{p}_{II})
}\,.
\end{equation}
Note that up to an overall prior on the model as a whole this is simply the ratio of the posterior probabilities of the two models given the data.  As such, in the absence of a clear prior preference, the Odds Ratio provides a straightforward assessment of their relative success.  Since this necessarily incorporates the assumed priors, and therefore penalizes the effort to model the inter-epoch closure phase variability in two ways: first by the inclusion of a larger parameter space volume and second by a prior penalty on large $\phi_E$.  Nevertheless, the Odds Ratio conclusively disfavors the analysis in which the closure phase variability is ignored.  

The magnitude of the $\phi_E$ required are listed in Table \ref{tab:chi2e} both for the most likely model and after imposing the prior on $\phi_E$ described in Section \ref{sec:CIO}.  The marginalized shifts, $\bar{\phi}_E$, are roughly normally distributed with a mean of $3.34^\circ$ and standard deviation of $3.41^\circ$ implying a small but significant net shift in the closure phase measurements, possibly implying systematic deviations from the underlying model.  Such a shift could in principle be caused by turbulent structures within the accretion flow that are not modeled here; typically orbiting compact emission features produce a net bias in closure phases on the California-Hawaii-Arizona triangle that depends on the orientation of the disk \citep[see, e.g., left column of Fig. 5 in][]{2009ApJ...695...59D}, though future work is required to assess the magnitude and direction of these biases.  The largest closure phase shifts occur for epochs 11, 12 and 17, all of which are clearly discrepant with the others in Figure \ref{fig:CPdist}.  In the latter two cases, however, the shift is less than $2\sigma_\phi=7.7^\circ$.  The inferred mean closure phase during epoch 11 is sensitive to the fringe search method employed, varying between $-0.3^\circ$ and $3.5^\circ$.  As a result it is consistent with being less discrepent than epochs 12 and 17.  For this reason we do not consider the apparently large shifts required for epoch 11 to be significant at this time, and thus all shifts are within $2\sigma_\phi$ of zero.

\subsection{Cross-Epoch Fit Consistency}

Having established that our RIAF model class provides an adequate description of the EHT data, we now turn to assessing the consistency among epochs.  With the exception of flare events, there is no reason for the underlying image morphology of an aligned RIAF to evolve across epochs.  Misaligned accretion flows can precess on a variety of timescales, leading to secular changes in the reconstructed source parameters.  However, within the context of the RIAF model class considered here, we expect that the black hole parameters that characterize the images will be fixed.  Thus, while necessary the existence of good fits for each epoch is insufficient -- we also require that the reconstructed parameters be consistent among all epochs.

Posterior probability distributions are obtained by first marginalizing over the epoch-specific flux renormalizations (assuming a flat prior, which is well justified within the small range of fluxes permitted) and closure phase shifts (assuming a Gaussian prior).  This may be done analytically, and results in the same flux normalization estimate \citep[see also Appendix \ref{app:CPmarg}]{2014ApJ...784....7B}.  We further assume a flat prior on spin magnitude and isotropic prior on the spin orientation.  The resulting three-dimensional posterior probability distribution is a function only of $a$, $\theta$, and $\xi$.

\begin{figure*}
\begin{center}
\includegraphics[width=\textwidth]{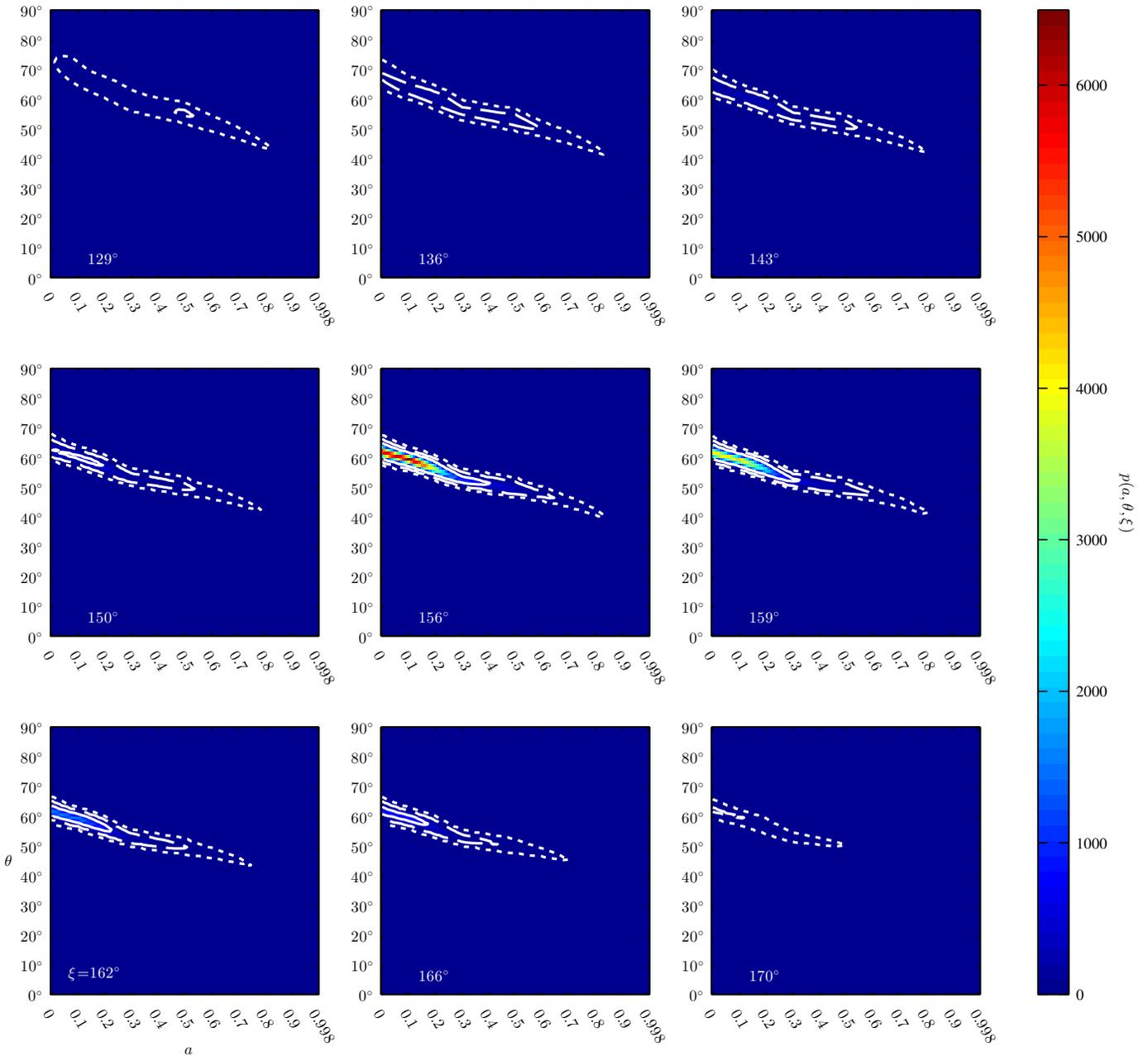}
\end{center}
\caption{Posterior probability distribution of the vector spin for the combined data set.  Each panel shows a slice of constant $\xi$, listed in the lower left corner.  The solid, dashed, and dotted contours show the 1$\sigma$, 2$\sigma$, and 3$\sigma$ confidence regions, respectively.}\label{fig:Pm3Datx} 
\end{figure*}

Example posterior probability distributions are shown in Figure \ref{fig:Pm3Depochs} for the combination of the visibility magnitude data (reproducing \citealt{2011ApJ...735..110B}) and three epochs of indicative closure phase data with samples from each year: epochs 5, 8, 16. All epochs have consistent solutions with the $\xi=156^\circ$ solution to the visibility magnitude fits.  Thus, as anticipated by the reasonable $\chi^2$ found for the combined fit in the previous section, a consistent set of spin parameters exist among all epochs.  This is shown for each fit parameter as a function of epoch in Figure \ref{fig:Pm1Depoch}.

This correspondence within epochs 1-4 has already been discussed in \citet{2011ApJ...735..110B}, and is to a significant degree a natural consequence of the consistency of the visibility amplitudes with each other.  A similar conclusion largely holds for epochs 5-17, where the consistency between the evolution of the closure phase evolutions implies that any satisfactory model will be similar on all epochs.  This is explicitly exhibited by the comparison of the best-fit model with the closure phase data in Figure \ref{fig:CPs}, which shows the same characteristically rising evolution.  Nevertheless, the data consistency is not absolute, as seen in Figure \ref{fig:Pm1Depoch}, where small variations do appear in the epoch-specific parameter reconstructions.

More remarkable is the consistency among qualitatively distinct classes of intereferometric data.  Earlier analyses have relied on the visibility magnitudes alone (resulting in the $180^\circ$ position angle degeneracy evident in Figures \ref{fig:Pm3Depochs} and \ref{fig:Pm1Depoch}).  While it may be reasonably expected that high-quality fits to subsequent, consistent visibility magnitude measurements will produce similar parameter estimates, this is certainly not true for visibility phases, and therefore closure phases.  A clear counter example is the statistically successful (though comparatively disfavored) Gaussian spot model for the emission region which predicts identically vanishing closure phases \citep[e.g.,][]{2011ApJ...735..110B}.  Thus, the recent set of closure phase measurements present an a priori test of the original visibility-magnitude selected RIAF models.  As a consequence, the consistency among epochs, and therefore consistency among classes of interferometric observables, is an impressive success of the RIAF picture.

\subsection{Black Hole Spin Estimation}

\begin{figure}
\begin{center}
\includegraphics[width=0.45\textwidth]{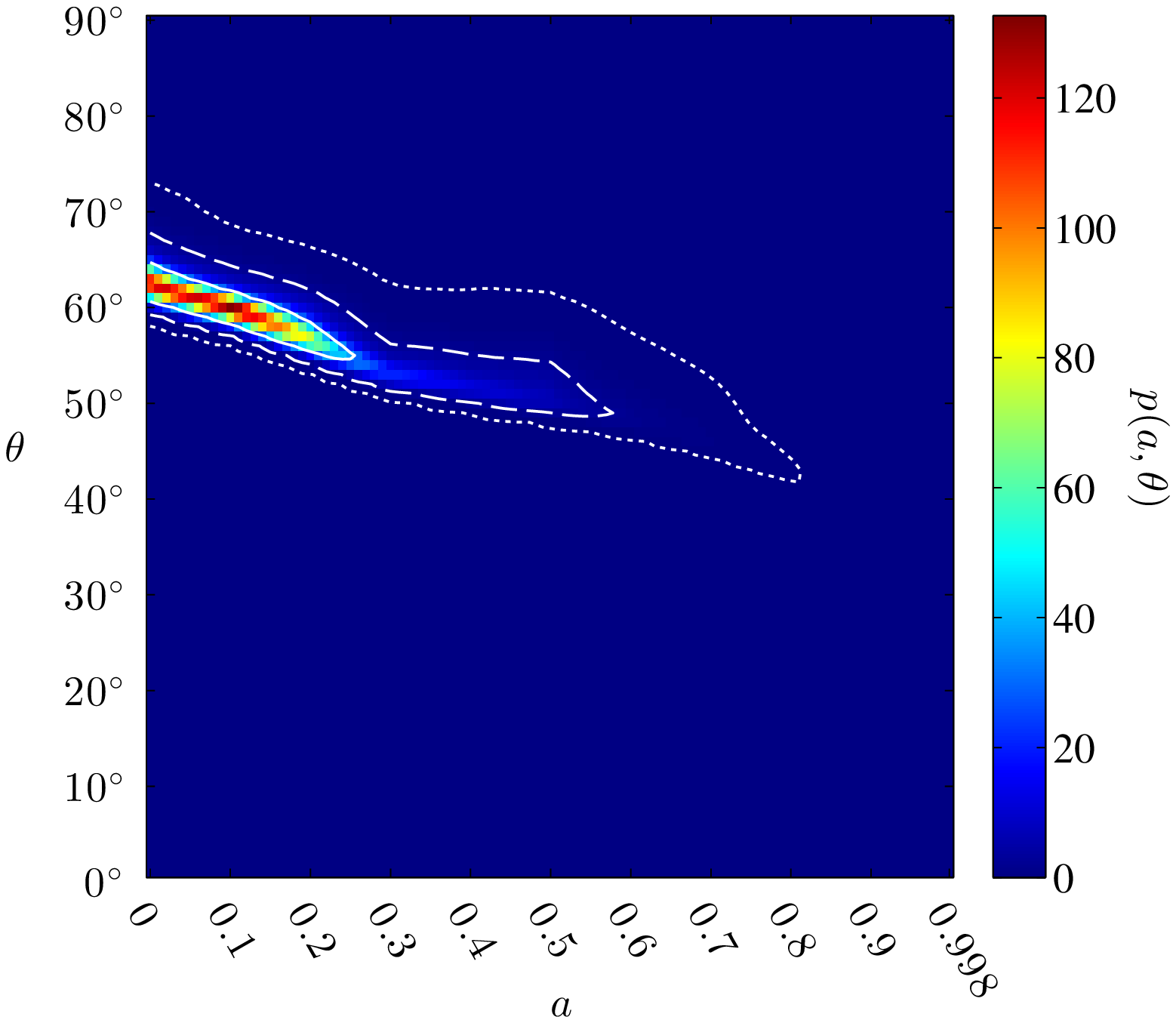}
\end{center}
\caption{Posterior probability of spin magnitude and inclination, marginalized over position angle.  Note that there is a reflection degeneracy in the reconstructed $\theta$.  Solid, dashed, and dotted contours show the 1$\sigma$, 2$\sigma$, and 3$\sigma$ confidence regions, respectively.}\label{fig:Pm2D}
\end{figure}

\begin{figure*}
\begin{center}
\includegraphics[width=0.32\textwidth]{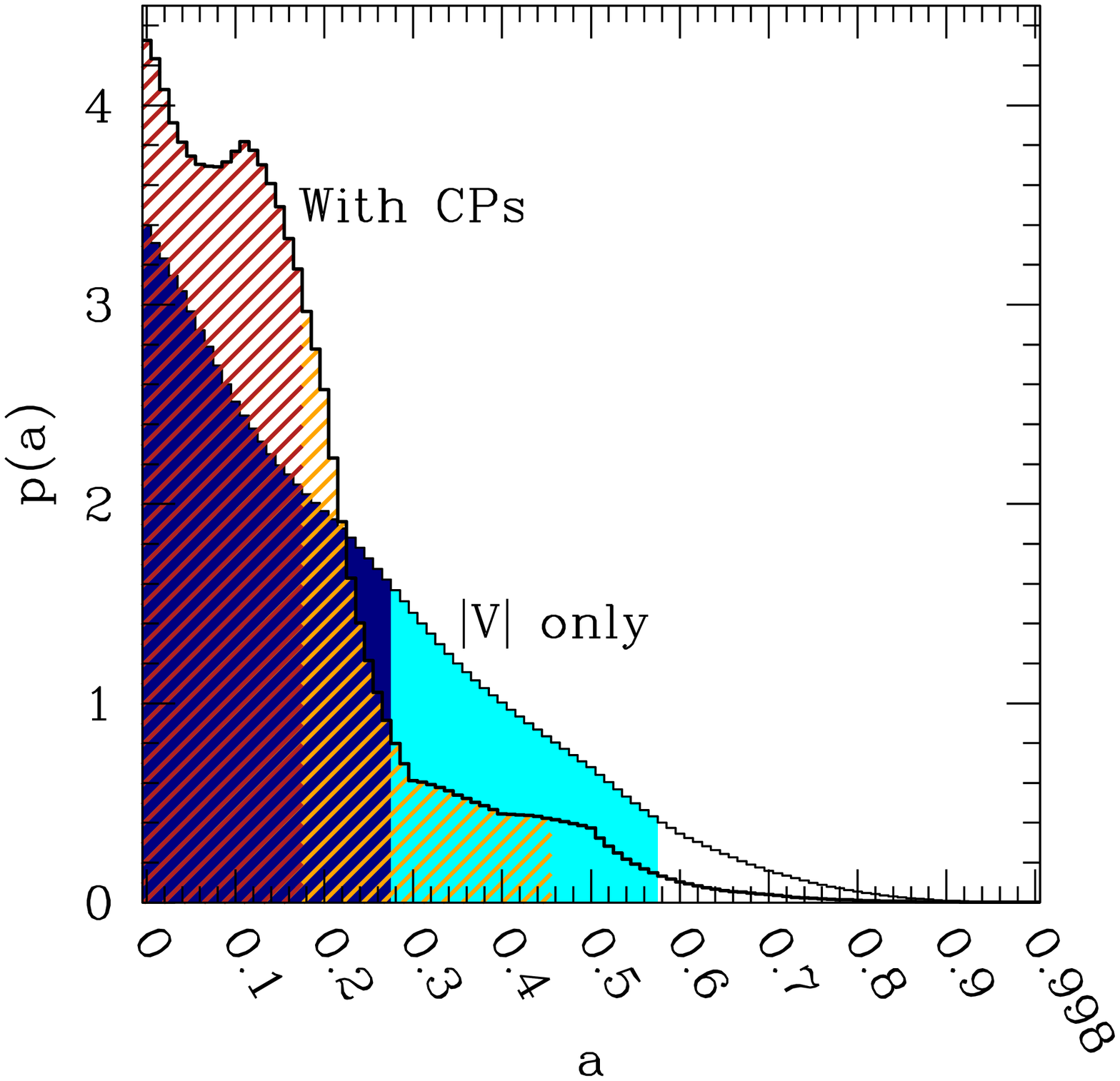}
\includegraphics[width=0.32\textwidth]{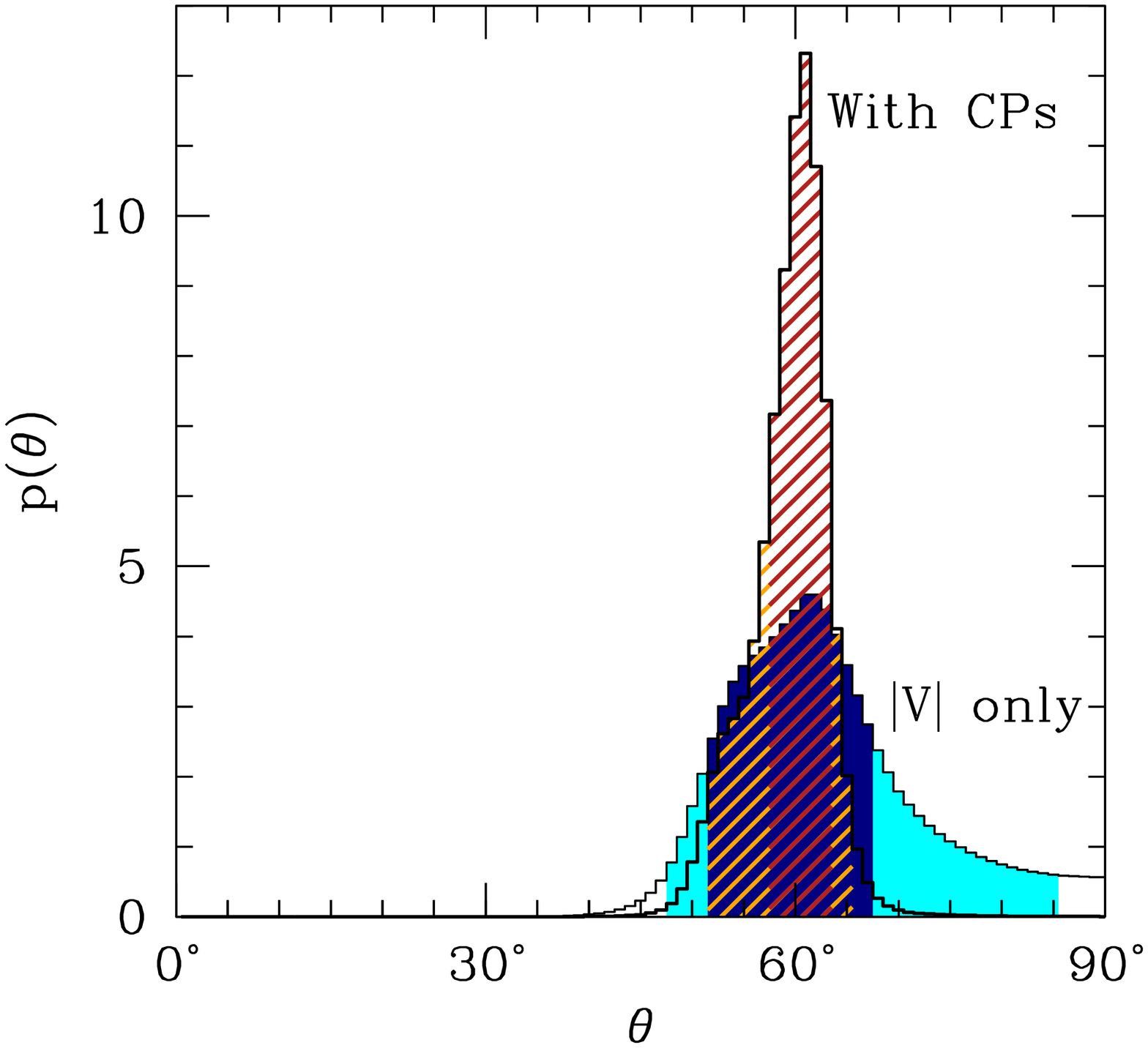}
\includegraphics[width=0.32\textwidth]{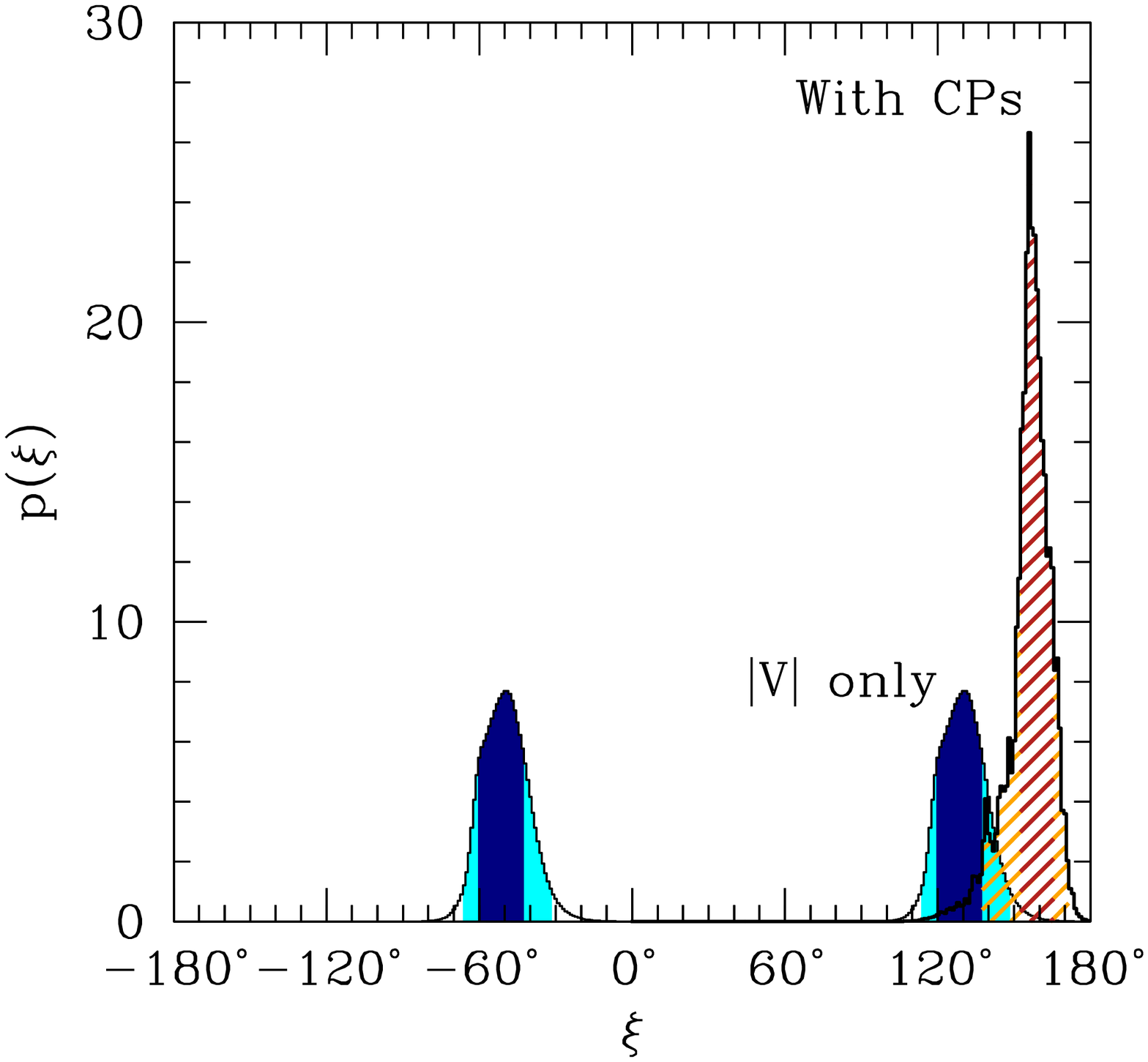}
\end{center}
\caption{Posterior probability as a function of spin magnitude (left), inclination (center), and position angle (right), marginalized over all other parameters.  For comparison the probability distributions obtained from the visibility magnitudes alone in \citet{2011ApJ...735..110B} are shown in blue.  The dark and light blue hatched regions show the 1$\sigma$ and 2$\sigma$ regions.}\label{fig:p1D}
\end{figure*}

The impact of black hole spin on the mm-wavelength images of RIAFs is nontrivial and arises from a number of sources, including modifications to the null geodesics traversed by the mm photons and the underlying dynamics of the emitting material.  Importantly, for RIAFs it is not possible to assign the impact of spin to modification of the ISCO, as is commonly the case for thin disks \citep[this is evident in efforts to constrain perturbations to Kerr with similar analyses, e.g.,][]{2014ApJ...784....7B}.  For the set of spectrally-fit, aligned RIAF models considered here the magnitude of the spin is determined primarily by the size of the emission region: higher spins result in both more rapidly orbiting material in the inner regions of the accretion flow and a greater degree of alignment between the emitting gas and mm-photon directions, leading to more severe Doppler boosting and beaming, all of which at {\em fixed} total flux produces a more compact image.  Because we consider an aligned class of RIAF models we necessarily mix the impact of the black hole spin and accretion flow inclination.  The degree to which one dominates the constraint over the other depends on the orientation and magnitude of the black hole spin; at vanishing spins the orientation reconstruction is essentially dominated by that of the accretion flow.

Having verified that high-quality fits exist and are consistent among epochs, we now turn to the estimation of the black hole spin magnitude and orientation.  The combined epoch fits are shown in Figure \ref{fig:Pm3Datx} with the two-dimensional probability distribution marginalized over position angle shown in Figure \ref{fig:Pm2D} and the one-dimensional marginalized probability distributions for each parameter shown in Figure \ref{fig:p1D}.  As found in \citet{2011ApJ...735..110B}, the allowed parameter space is restricted to a narrow region, being primarily degenerate in $a$ and $\theta$.  Unlike \citet{2011ApJ...735..110B}, the allowed region is now becoming constrained in all directions, with the result that the three-dimensional and one-dimensional parameter estimations are becoming similar.   That is, the large-scale correlations that have complicated the characterization of the allowed spin parameters are disappearing.

\subsubsection{Spin Orientation}
For the quiescent image models considered here, an approximate reflection symmetry across the equatorial plane of the black hole, broken only by absorption by the black hole itself and the non-trivial optical depth of the accretion flow, prevents distinguishing between inclinations with the same $|\cos\theta|$, i.e., between $\theta$ and $180^\circ-\theta$.  The images produced by the two cases are related by an image reflection, and thus we independently fit model libraries for both $\theta\le90^\circ$ and $\theta\ge90^\circ$.  At present the two model classes are indistinguishable, both having nearly equally good fits and having an Odds Ratio of unity (see Table \ref{tab:chi2}).  Figures \ref{fig:Pm3Depochs}--\ref{fig:Pm2D} show the case when $\theta\le90^\circ$, illustrative of both.  Figure \ref{fig:p1D} show results marginalized over both potential inclinations.  Note that dynamical observations within the context of shearing turbulent flows can explicitly break this degeneracy \citep{2015arXiv150507870J}.

Nevertheless, despite the degeneracy, the quantitative estimates for the inclination are significantly improved, with constraints that are nearly a factor of two better than those in \citet{2011ApJ...735..110B}: $\theta={60^\circ}^{+5^\circ+10^\circ}_{-8^\circ-13^\circ}$.\footnote{In parameter estimates the subscripts and superscripts indicate the 1$\sigma$ and 2$\sigma$ errors in each direction.}  When marginalized over all other parameters, this results in an inclination estimate of $\theta={60^\circ}^{+2^\circ+5^\circ}_{-4^\circ-8^\circ}$, again more than a factor of two improvement over the marginalized estimates of \citet{2011ApJ...735..110B}.  In both cases these are consistent at the 1$\sigma$ level with the previous constraint, as shown explicitly in the center panel of Figure \ref{fig:p1D}.

Unlike the visibility magnitudes, closure phases are able to conclusively select a single position angle, marking both a qualitative and quantitative improvement in its determination.  The value selected is $\xi={156^\circ}^{+10^\circ+14^\circ}_{-17^\circ-27^\circ}$ east of north.  After marginalizing over spin magnitude and inclination the position angle estimate is $\xi={156^\circ}^{+10^\circ+16^\circ}_{-3^\circ-18^\circ}$, consistent with the values reported in \citet{2011ApJ...735..110B} at between the 1$\sigma$ and 2$\sigma$ level.  

The small shift in the reconstructed position angle may indicate a mild tension between the orientations inferred from the visibility magnitudes alone and those from the closure phases alone.  The origin of this tension remains unclear, though it may derive from a variety of sources.  Much of the angular information in the visibility magnitude analyses is found in the early and late visibility measurements along the long Hawaii-CARMA and Hawaii-SMT baselines, where Sgr~A*'s elevation at all stations is at its lowest and therefore the flux calibration most challenging.  Errors in the reconstructed station fluxes would mimic structure in the north-south direction, modifying the reconstructed image orientation.  These calibration difficulties do not impact the inferred structure from closure phase measurements, which derive instead from their non-zero value at all times.  Alternatively, short time variability associated with compact features in the accretion flow can induce significant dynamical deviations in the closure phases that are not accounted for in our attempts to address the inter-epoch closure phase variability \citep{2009ApJ...695...59D}.  As with calibration uncertainties, this will modify the position angle most strongly as a result of small north-south baselines at present.  Therefore, unsurprisingly, the imminent improvements in the north-south coverage of the EHT will be critical in addressing the position angle of Sgr A* and the attendant implications for accretion modeling.

\begin{figure*}
\begin{center}
\includegraphics[width=\textwidth]{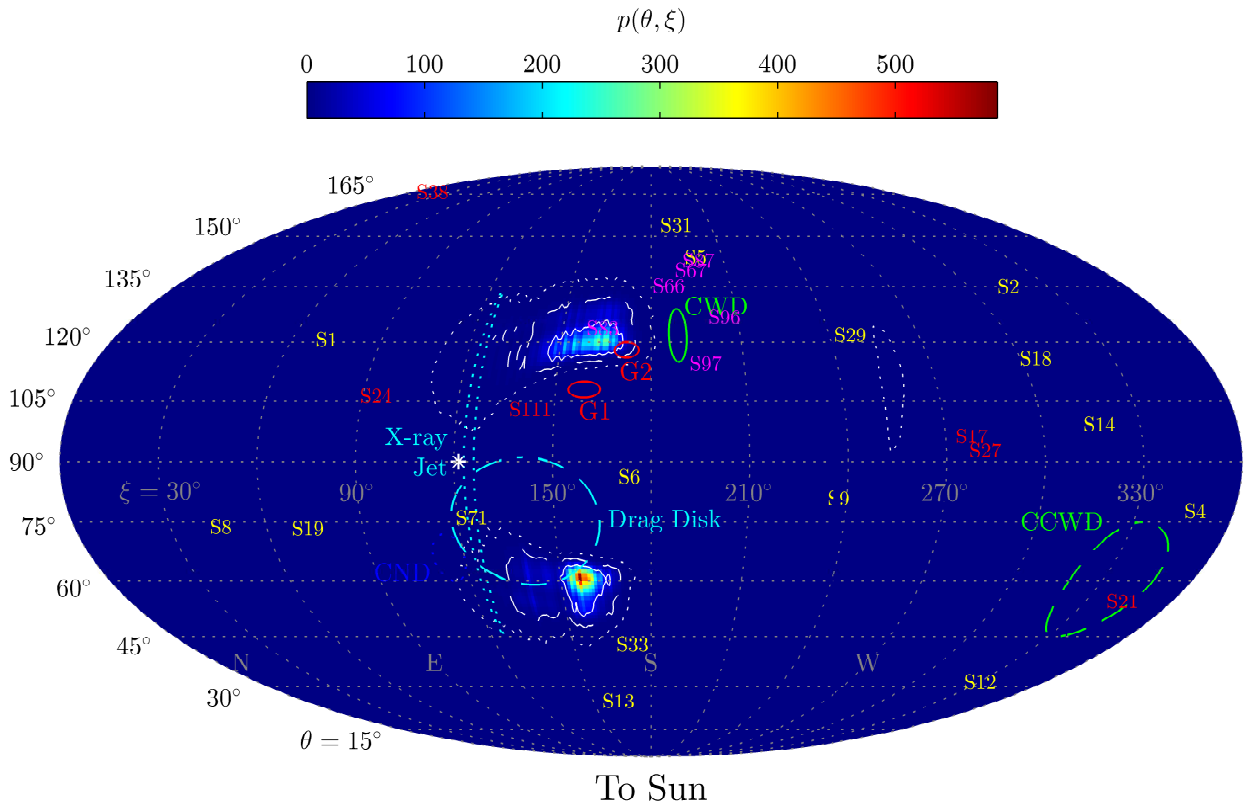}
\end{center}
\caption{Orientation of the spin of Sgr~A*, marginalized over spin
  magnitude, in comparison to the angular momentum vectors of other
  features in the Galactic center.  The white solid, dashed, and
  dotted contours show the 1$\sigma$, 2$\sigma$, and 3$\sigma$
  confidence regions, respectively; the inclination degeneracy is
  manifest in the two islands at $\theta=60^\circ$ and $120^\circ$.
  For comparison, the angular momenta of the clockwise (CWD) and
  counter-clockwise stellar disks (CCWD), infrared gas clouds G1 and
  G2, a handful of the S-stars.  The S-star colors indicate the
  stellar type: yellow and red are early-type and late-type B stars,
  while magenta are Wolf-Rayet stars associated with the CWD.  Also
  shown are the orientations of the accretion flow as reconstructed from its assumed impact on G1 and G2 (Drag Disk), putative X-ray jet feature,
  circum-nuclear disk (CND), and Galactic rotation axis
  (\ding{107}).}\label{fig:orientation}
\end{figure*}

The reconstructed orientation of Sgr~A* is in remarkable agreement with a variety of features within the Galactic center \citep{2015ApJ...798...15P}, shown in Figure \ref{fig:orientation}.  Most notably, one of the inclination solutions is aligned at the 1$\sigma$ level with the inferred orbit of the infrared-luminous gas clouds G1 and G2 \citep{2015ApJ...798..111P}.  Equally suggestive is the near-alignment with the clockwise disk of young stars (CWD), believed to be responsible for feeding the accretion flow onto Sgr~A*  \citep{2009ApJ...690.1463L,2010RvMP...82.3121G}.  These are natural for two reasons, relating either to the structure of the accretion flow or the spin of the black hole.  

First, the strong viscous coupling within RIAFs, arising from the large-scale magnetic fields that result from their large scale heights, prevents the Bardeen-Petterson effect from efficiently aligning the disk with the black hole spin \citep{1975ApJ...195L..65B, 2007ApJ...668..417F}.  Thus, for small black hole spins the orientation of a RIAF is determined by the original angular momentum of the accreting gas.\footnote{For high black hole spins the accretion flow may precess \citep{2005ApJ...623..347F}, form unstable accretion streams and shocks \citep{2007ApJ...668..417F, 2011ApJ...730...36D}, or break \citep{2012ApJ...757L..24N}.}

Second, if these stars formed in situ, the accretion of the associated gas disk is sufficient to reorient the black hole spin, naturally accounting for the inferred alignment.  To overcome the local tidal forces requires a disk mass of $M_{\rm disk}\approx10^4$--$10^5\,M_\odot$ extending to a distance of $R_{\rm disk}\approx0.4~\pc$, of which only roughly $10^3\,M_\odot$ would have resulted in stars, the remainder accreting onto Sgr~A* over the past $10^6~\yr$ \citep{2007MNRAS.374..515L,2009ApJ...697.1741B}.  The orbital angular momentum in this gas is roughly
\begin{equation}
J_{\rm gas} 
\approx 
\sqrt{GM R_{\rm disk}} M_{\rm disk}
=
\frac{J_{\bullet}}{a} \sqrt{\frac{R_{\rm disk} c^2}{GM}} \frac{M_{\rm disk}}{M}
\gtrsim
3 \frac{J_{\bullet}}{a}
\end{equation}
where $J_{\bullet}$ is the angular momentum of the black hole.  Thus generally there is enough angular momentum in the associated accretion disk to align the black hole, the two can be efficiently coupled.  This depends, in turn, on the details of the disk accretion.

The accretion rate at the black hole is highly uncertain, depending critically on the impact of disk winds.  For wind models similar to those in the models we have considered here, $\dot{M}(r)\propto r^{0.45}$, and thus the accretion rate at the horizon is only $0.15\%$ of that at in the star forming region.  Nevertheless, even in the absence of wind losses, from \citet{2007MNRAS.374..515L} the accretion rate near $r=R_{\rm disk}$ is $\dot{M}_{\rm disk} \approx 0.02 \dot{M}_{\rm Edd}$ assuming typical values.  Thus in the presence of even marginal mass loss the accretion will proceed within the RIAF regime.  As a consequence, magnetic torques are expected to strongly couple the accretion flow over the large distances required to align the black hole spin.  

It is noteworthy that the implied total energy output, $3\times10^{53}~\erg\,\s^{-1}$--$2\times10^{56}~\erg\,\s^{-1}$ (assuming radiative efficiency of 1\%) is comparable the energy budget required to inflate the Fermi bubbles, kpc-scale gamma-ray features above and below the Galactic plane believed to have occurred contemporaneously with the formation of the CWD.  This is consistent with the emerging picture of Sgr~A*'s recent accretion history, characterized by the recent end of a more active phase roughly $10^6~\yr$ ago 
\citep[see, e.g.,][and references therein]{2014IAUS..303..333P}.

\begin{figure*}
\begin{center}
\begin{tabular}{ccc}
\includegraphics[width=0.32\textwidth]{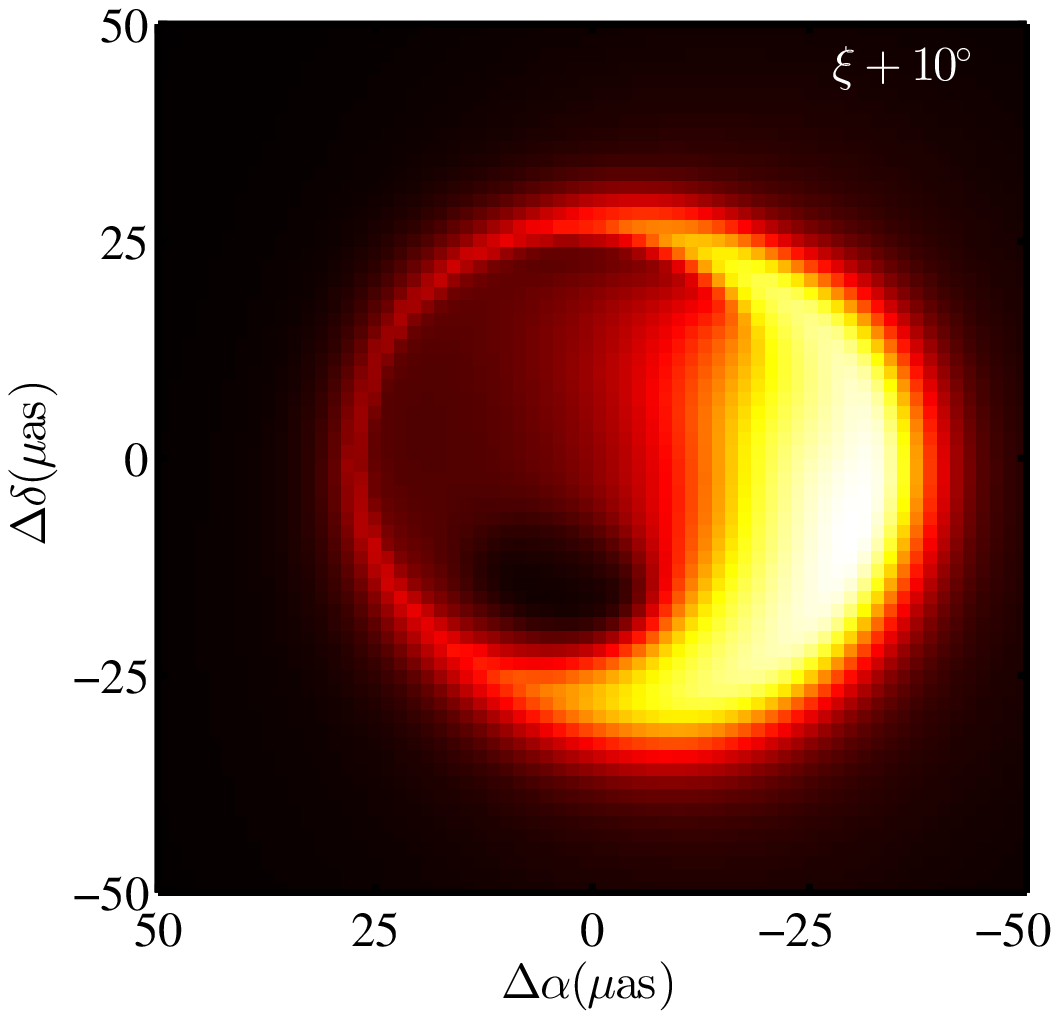} &
\includegraphics[width=0.32\textwidth]{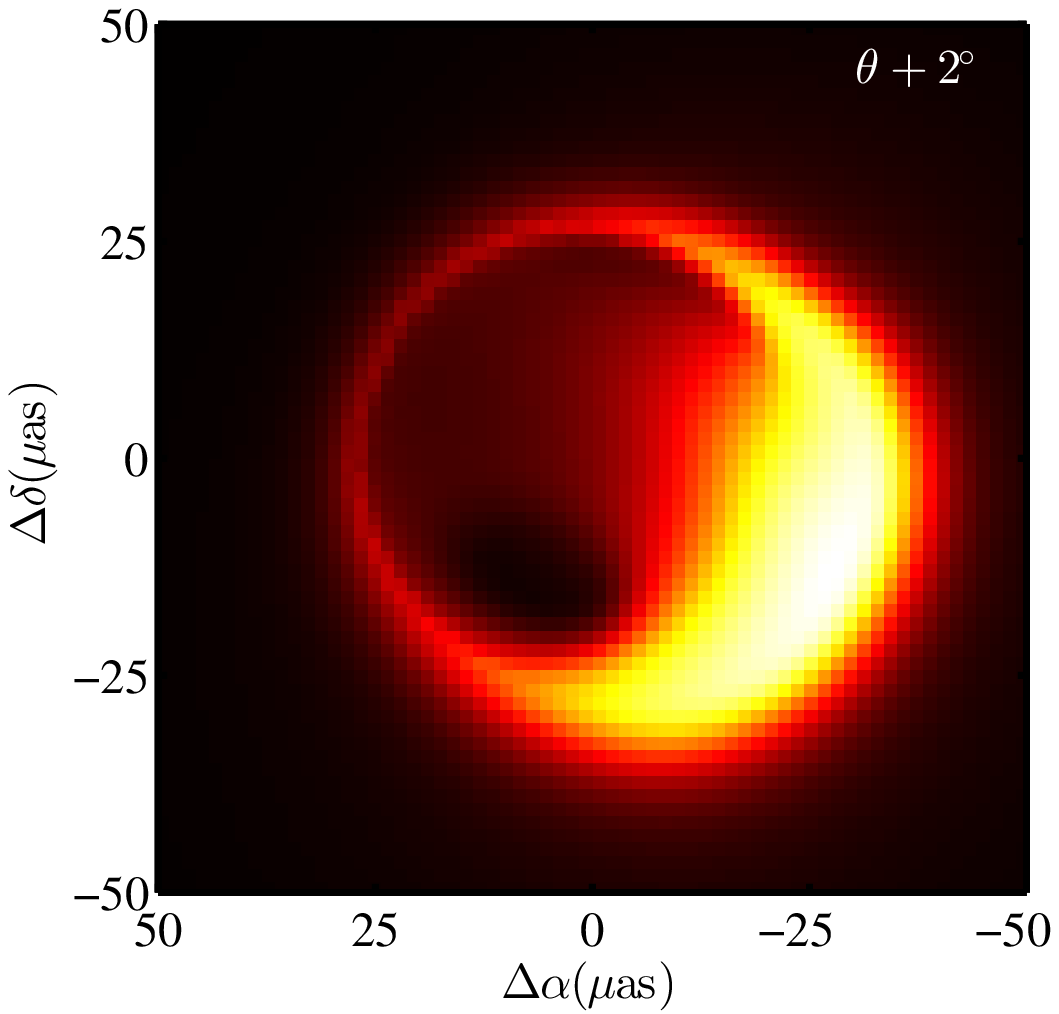} & \\
\includegraphics[width=0.32\textwidth]{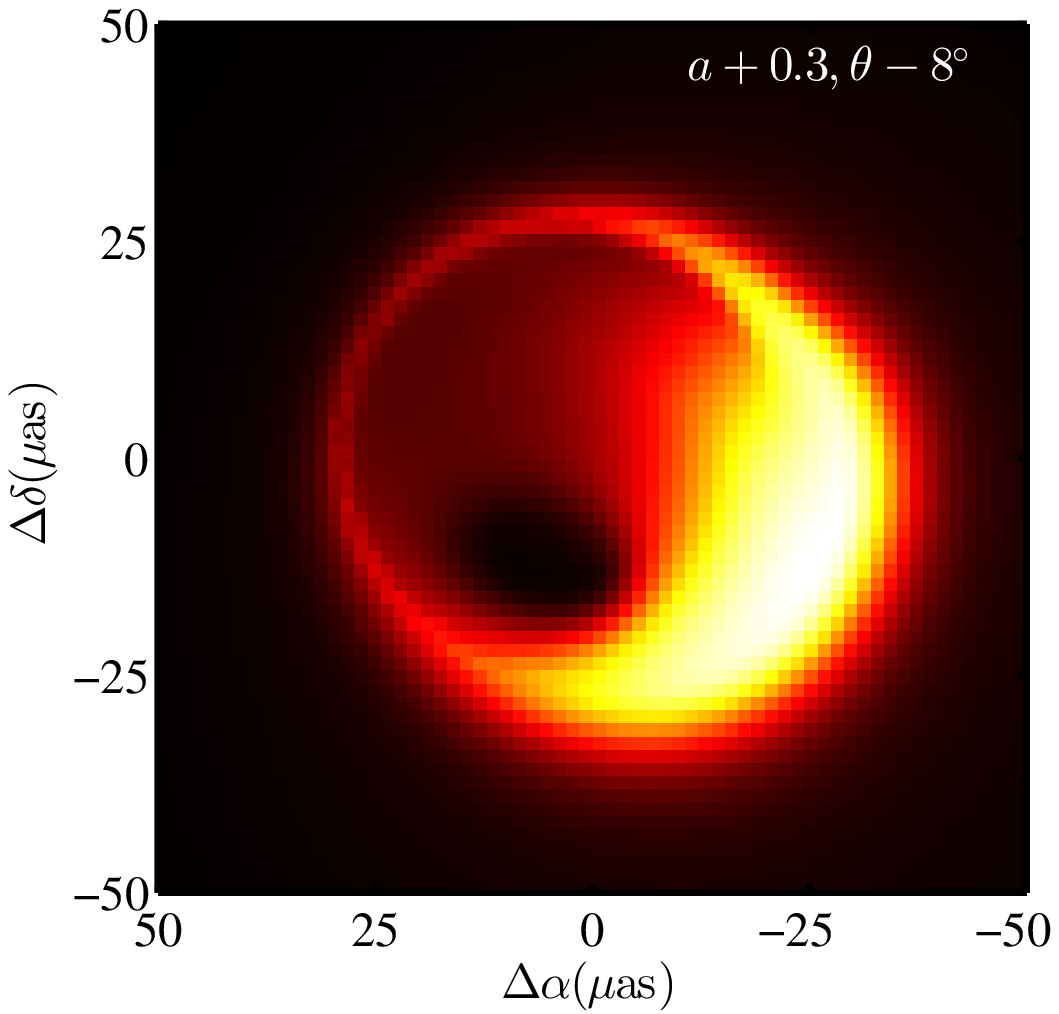} &
\includegraphics[width=0.32\textwidth]{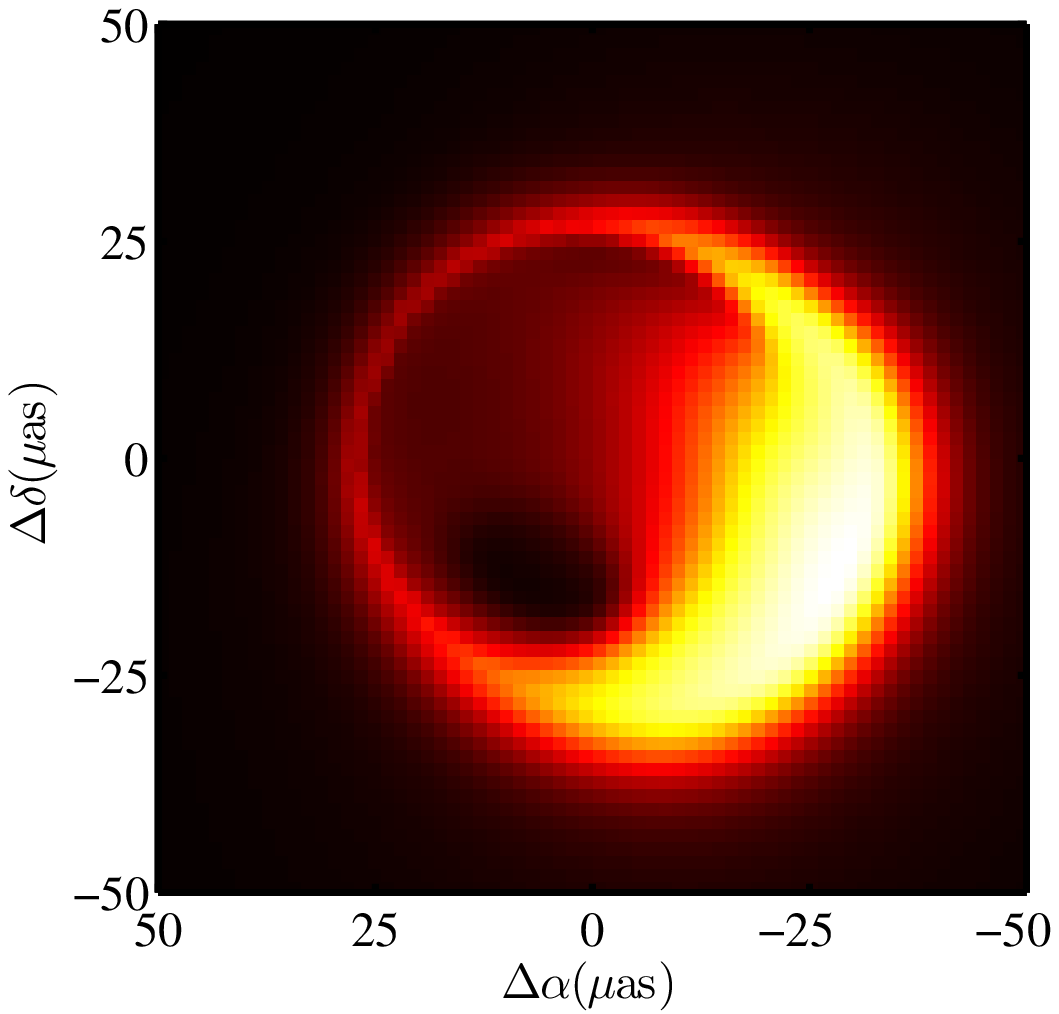} &
\includegraphics[width=0.32\textwidth]{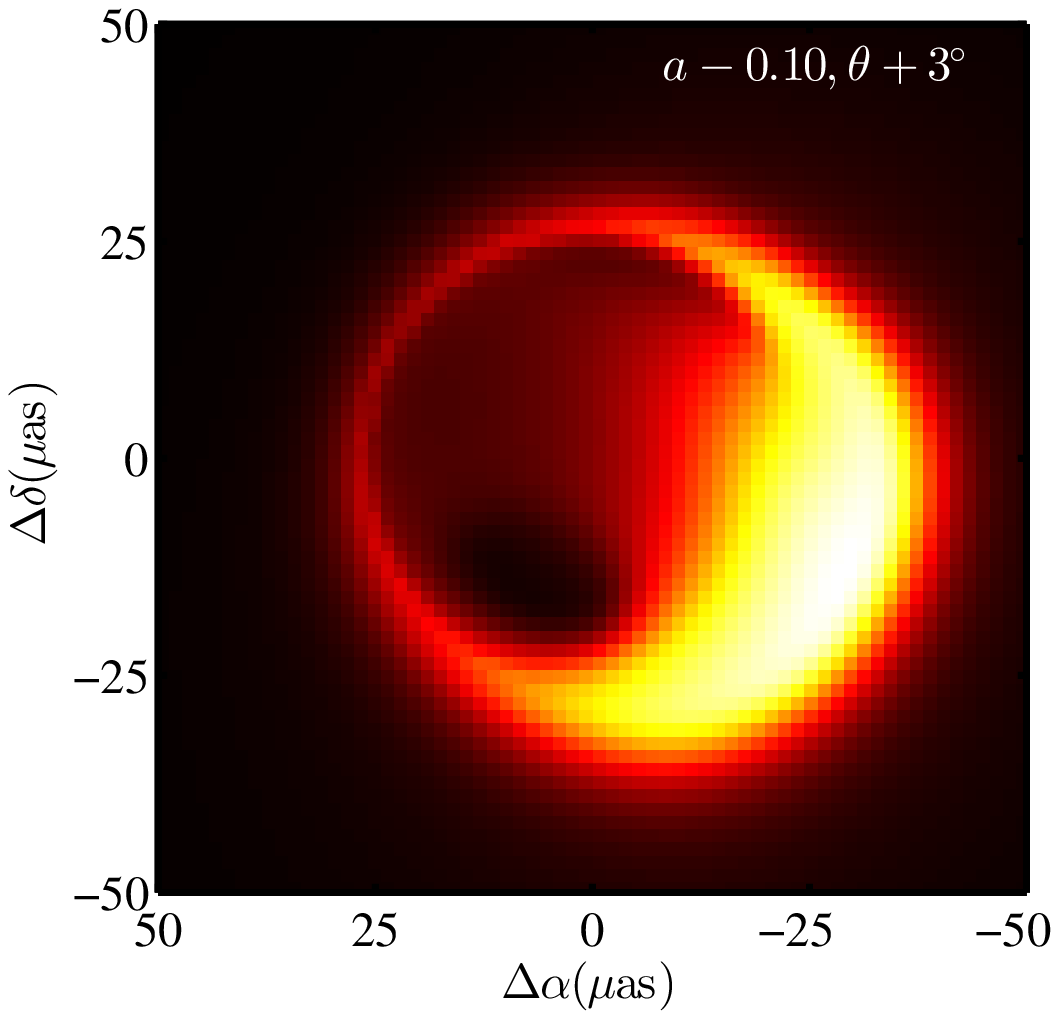} \\
&
\includegraphics[width=0.32\textwidth]{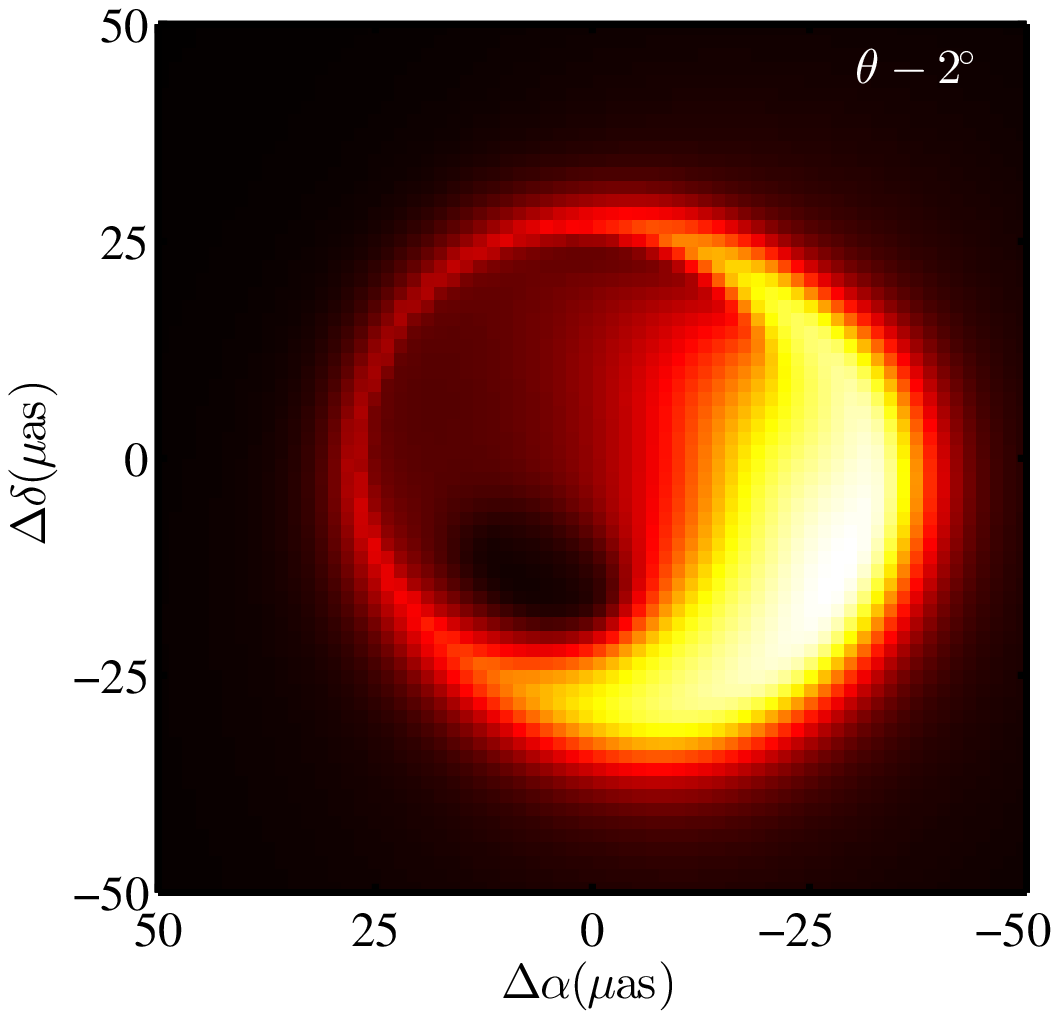} &
\includegraphics[width=0.32\textwidth]{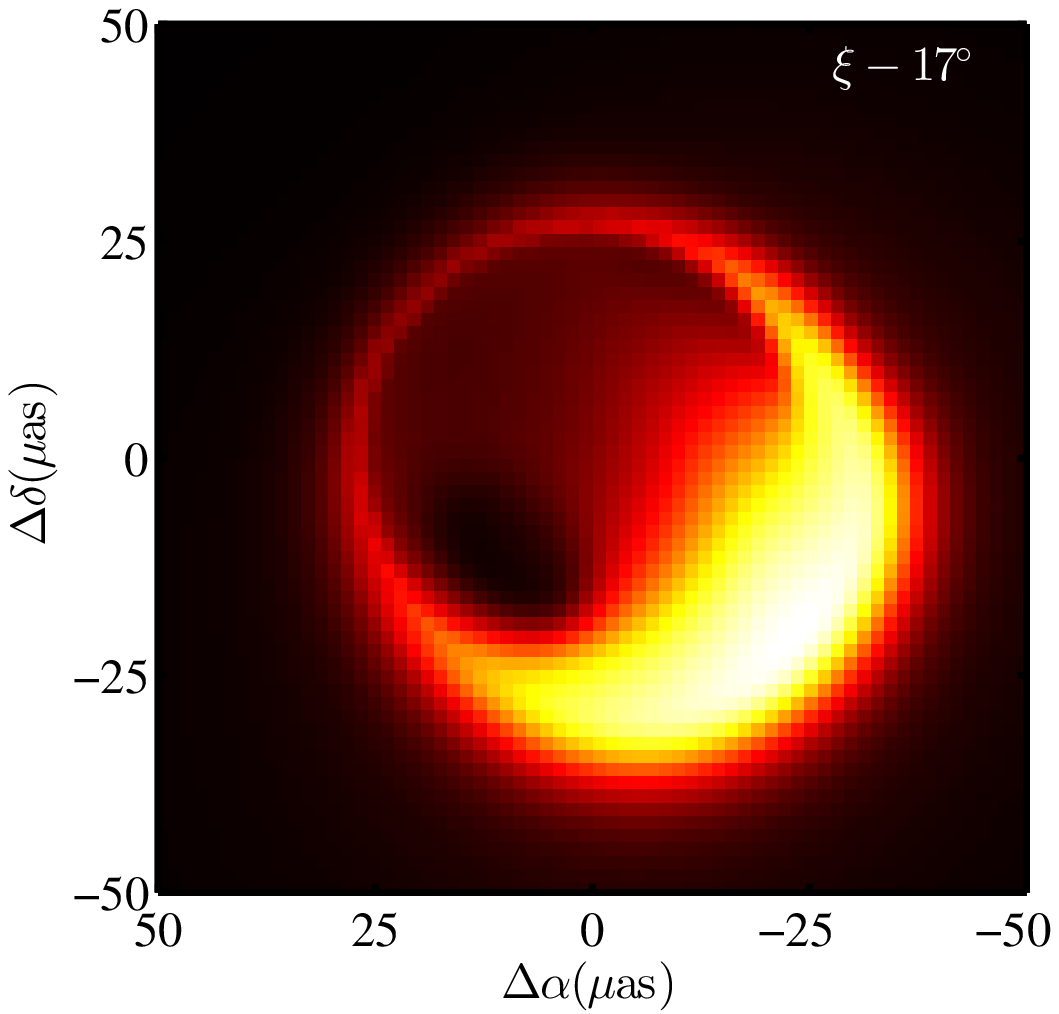}
\end{tabular}
\end{center}
\caption{Images associated with the most probable model ($a=0.10$, $\theta=60^\circ$, $\xi=156^\circ$; center) and 1$\sigma$ deviations in the best fit parameters.  The shifts are indicated in the upper right corner of each image.  In all the intensity scale is linear and consistently normalized to that of the center image.}\label{fig:images}
\end{figure*}

The reconstructed orientation of Sgr~A* is potentially in alignment with the X-ray jet feature reported by \citep{Li-Morr-Baga:13}.  In addition, it is consistent with the orientation of the large-scale accretion flow as inferred from dynamical drag required to place G1 and G2 on similar orbits \citep{2016MNRAS.455.2187M,2016arXiv160202760M}.  However, either would necessarily preclude the alignment with the clockwise disk.  Notably, {\em no} solution for the orientation of Sgr~A* aligns with the more distant counter-clockwise disk of stars (CCWD), circum-nuclear disk at 3~pc (CND), Galactic rotation axis \citep[and thus kpc-scale Fermi bubbles,][]{2010ApJ...724.1044S,2014ApJ...793...64A}, or the nearby S-stars \citep{2009ApJ...692.1075G,2010RvMP...82.3121G}.

\subsubsection{Spin Magnitude}

The magnitude of Sgr~A*'s spin remains tightly correlated with inclination, as clearly seen in Figures \ref{fig:Pm3Datx} and \ref{fig:Pm2D}.  This is marginally weaker than found in \citet{2011ApJ...735..110B}, with $a$ and $\theta$ related by
\begin{equation}
\theta \approx 63^\circ - 31^\circ a \pm 1.5^\circ \pm 3.5^\circ\,.
\end{equation}
The spin magnitude is not significantly correlated with position angle.
Consequently, the improvements of the constraints on the spin magnitude mirror those for the spin inclination.  Unlike \citet{2011ApJ...735..110B} the most probable spin magnitude, $a=0.10^{+0.30+0.56}_{-0.10-0.10}$, is formally nonzero, though it remains consistent with $a=0$, in which case the orientation corresponds to that of the accretion flow angular momentum.  The most probable model is shown in Figure \ref{fig:images}.

Marginalizing over orientation produces strictly upper limits of $a=0^{+0.18+0.45}$, ignoring the possibility of anti-alignment (corresponding here to $a<0$), and shown in Figure \ref{fig:p1D}.  As before, the resulting spin magnitude estimates are consistent at the 1$\sigma$ level with those in \citet{2011ApJ...735..110B}.  This is complicated, however, by the choice of spin magnitude prior.  Regardless, the conclusion in \citet{2011ApJ...735..110B} that the spin magnitude must be small continues to be supported by the subsequent EHT observations. 

For clarity we have adopted a flat prior.  While ideally a prior may be obtained from simulations of supermassive black hole growth \citep[e.g.,][]{2005ApJ...620...69V, 2013ApJ...775...94V, 2012MNRAS.423.2533B}, considerable astrophysical uncertainties persist preventing this in practice.  A completely agnostic spin magnitude prior would favor high spins and is $\propto a^2$, corresponding to the volume of the angular momentum phase space of a spinning top.  While the value of such an arbitrary prior is questionable, even in light of its emphasis on high spins $a<0.51$ at 2$\sigma$ and $a<0.73$ at 3$\sigma$.  Thus, regardless of the prior adopted, within the context of the RIAF models considered here very high spins are excluded at high significance.

\section{Conclusions} \label{sec:conclusions}

Models for the radio emission from Sgr~A* based on the RIAF paradigm continue to provide an excellent description of the horizon-scale millimeter-wavelength structure as probed by the EHT.  This is consistent across sixteen observation epochs, extending over seven years.

This consistency is nontrivial; twelve of the new observation epochs consist of closure phase observations providing an independent test of the RIAF picture.  There is no a priori reason to have expected RIAF models fit to the earlier visibility magnitude data to continue to provide a satisfactory fit to the subsequent closure phase data.  For example, the previously successful Gaussian models are incapable of reproducing the non-trivial and time-varying closure phases observed.  Thus, these later epochs have provided a critical a priori test of the RIAF picture.

With the advent of a large number of closure phase measurements it has become clear that it is necessary to model the inter-epoch variations associated with small-scale structure within the image.  Possible sources of these features include intrinsic accretion flow structure, e.g., that arising from the magnetohydrodynamic turbulence implicated in the transport of angular momentum, and refractive structures in the intervening scattering screen.  Thus it appears that the inter-epoch closure phase fluctuations will provide a means to probe at least two classes of important dynamical features for imaging Sgr~A*.

The resulting constraints on the black hole spin orientation and magnitude have been improved by roughly a factor of two over those presented in \citet{2011ApJ...735..110B}.  This includes a substantial qualitative improvement: the previous $180^\circ$ rotational degeneracy has now been conclusively broken, yielding a single position angle solution.  Nevertheless a mild tension (1$\sigma$) between the position angles inferred from the visibility magnitudes and closure phases may indicate a remaining unmodeled systematic.  One of the two potential reconstructed orientations is in remarkable agreement with the orbits of the infrared gas clouds G1 and G2, as well as the clockwise disk of stars that is believed to source Sgr~A*'s accretion flow.  This suggestive association supports this conclusion, though additional observations will be required to break the remaining reflection degeneracy in the inclination.

The magnitude of the black hole spin continues to be consistent with zero, though values as high as $a=0.45$ remain possible at the 2$\sigma$ level.  Regardless of the assumed spin magnitude prior, high values ($a>0.73$) are excluded at high confidence within the class of RIAF models considered here.  Small values of black hole spin obtain further weak circumstantial support from the apparent alignment between the spin and the angular momentum of the presumed origins of Sgr A*'s accretion flow, the CWD and infrared gas clouds G1 and G2, all of which are natural only when the spin is small.  If true this would partially explain the lack of a vigorous jet in the Galactic center.

\begin{appendix}

\section{Structure Driven Closure Phase Variability} \label{app:CPvar}

Assessing the impact of small-scale structure on interferometric observables is complicated by the variety of dynamical timescales and potential structural correlations in the emission region.  Some of these derive from the source of variability while others are imposed externally.  As such, despite its great potential value as a means to study the dynamics of both the intrinsic emission region and subsequent propagation effects, a definitive treatment of the variability in closure phases lies well beyond the scope of the present study.  Nevertheless, here we make an attempt to roughly quantify the magnitude of the impact scattering or accretion flow turbulence can have on the observed closure phases, motivating the size of the closure phase shifts described in Section \ref{sec:CPFs} within a physical context.

The comparative shortness of the CARMA-SMT baseline to the Hawaii baselines, and thus the narrow nature of the triangle on which the closure phases are defined, admits a simple interpretation of the relationship between the underlying variable image structure and the variability of the resulting closure phase.  Here we derive this relationship and demonstrate that the observed degree of closure phase variability can be explained by inter-epoch variable structures on the angular scales probed by the long baselines of order 10\%.

We begin by obtaining an expression for the closure phase associated with a highly anisotropic triangle, i.e., in the squeezed-triangle limit.\footnote{This is one of the limits often employed to study the impact of gravitational lensing on the cosmic microwave background \citep{2003JHEP...05..013M,2004PhRvD..70h3532C,PhysRevD.80.043510}.  Unlike the CMB, however, Sgr A* is not statistically isotropic, and thus the orientation of the triangle is of critical importance.}  To do this we specify long and short baselines $\u$ and $\du$, respectively, with the hierarchy $\u\gg\du$.  In terms of these the baselines connecting the three stations are:
\begin{equation}
  \u_1 = \u - \frac{\du}{2}\,,\quad
  \u_2 = -\u - \frac{\du}{2}\,,\quad\text{and}\quad
  \u_3 = \du\,.
\end{equation}
Note that these necessarily satisfy the closure relation (hence the need for only two baselines to define the closure triangle).  The bispectrum is then given in terms of the complex visibilities by
\begin{equation}
\begin{aligned}
B_{123}
&=
V\left(\u-\frac{\du}{2}\right)
V\left(-\u-\frac{\du}{2}\right)
V\left(\du\right)\\
&=
V_0|V_u|^2 \left[ 1 + \bmath{M}\cdot\du + \mathcal{O}(\delta u^2) \right]
\end{aligned}
\end{equation}
where
\begin{equation}
\bmath{M}
=
(\grad_{\bf u} \ln V)_0
-
\frac{1}{2} \left[ (\grad_{\bf u} \ln V)_u - (\grad_{\bf u} \ln V^*)_u \right]
\end{equation}
in which we have used the identity $V(-\u)=V^*(\u)$.  The closure phase is the argument of the bispectrum, and thus
\begin{equation}
\Phi_{123}
= \Im\left(\ln B_{123}\right)
\approx \Im\left(\bmath{M}\cdot\du\right)\,,
\label{eq:Phi123}
\end{equation}
where we have again made use of the smallness of $\du$ and employed the fact that $V_0|V_u|^2$ is real.  Note that the closure phases vanishes when $\du=0$, as expected, and that $\bmath{M}$ is invariant to shifts in the image centroid (i.e., $V(\u)\rightarrow e^{2\pi i{\bf \Delta x}\cdot{\bf u}} V(\u)$) as required.  More importantly, the intrinsic variability in the closure phase is characterized entirely by the variations in $\bmath{M}$.

We begin by assuming an underlying quiescent structure with small fluctuations superposed, i.e.,
\begin{equation}
I(\bmath{x},t) = \bar{I}(\bmath{x})\left[1 + \Delta(\bmath{x},t)\right]\,,
\end{equation}
where $\Delta(\bmath{x},t)\ll1$.  Then, 
\begin{equation}
  V(\u,t)
  =
  \bar{V}(\u)\left[1 + \frac{\tilde{\Delta}(\u,t)*\bar{V}(\u)}{\bar{V}(\u)}\right]
  \equiv
  \bar{V}(\u) T(\u,t)\,.
\end{equation}
It is no longer true that the latter term in the braces is small.  Nevertheless, inserting this into Equation (\ref{eq:Phi123}) yields
\begin{equation}
\Phi_{123}
\approx
\bar{\Phi}_{123}
+
\Im\left[
(\grad_{\bf u} \ln T)_0
-
(\grad_{\bf u} \ln T)_u
\right]\cdot\du\,.
\end{equation}
Since the closure phase is insensitive to an over-all phase shift, we can set $\Im[(\grad\ln T)_0]=0$ identically, and hence
\begin{equation}
  \delta\Phi_{123}\equiv\Phi_{123}-\bar{\Phi}_{123} = -\Im\left[(\grad_{\bf u} \ln T)_u\right]\cdot\du\,.
  \label{eq:CPvar}
\end{equation}
That is, in this scenario the inter-epoch variability in the closure phase is determined by the time-variable structure in the transfer function on the angular scales associated with the long baseline.  The detailed structure of $\delta T(\u,t)$ depends on the underlying physical origin of the small-scale, time-variable structure of the image.  We consider two examples here: that due to refraction in the interstellar electron scattering screen, and structure resulting from turbulence in the accretion flow.

\subsection{Perturbative Weak Refractive Scattering}
Typical refractive angles are expected to be of order $20~\muas$, comparable to the scattering kernel width at $1.3~\mm$.  This is similar to the scale of small structures in the image.  Nevertheless, it is instructive to consider the weak refraction limit, i.e., where the image distortions due to scattering occur over scales small in comparison to the typical structures in the image.  We treat the strong scattering limit numerical in Section \ref{sec:strong_scatt}.

In the small-angle scattering limit, a refractive screen modifies the image via a distortion field $\bmath{\xi}(\bmath{x},t)$ which varies as the screen passes across the source.  That is, the observed intensity map is given in terms of the a fixed intrinsic image $\bar{I}(\bmath{x})$ by
\begin{equation}
  I(\bmath{x},t)
  =
  \bar{I}[\bmath{x} + \bmath{\xi}(\bmath{x},t)]
  \approx
  \bar{I}(\bmath{x}) \left[ 1 + \bmath{\xi}(\bmath{x},t)\cdot\frac{\grad \bar{I}(\bmath{x})}{\bar{I}(\bmath{x})}\right]\,,
\label{eq:Iwr}
\end{equation}
i.e., $\Delta = (\xi\cdot\grad\bar{I})/\bar{I}$.  The resulting expression for $T(\u,t)$ is then given by
\begin{equation}
  T(\u,t) = 1 + \frac{1}{\bar{V}(\u)} \left[\tilde{\bmath{\xi}} * \left(\frac{2\pi i \u}{\lambda} \bar{V}\right) \right](\u,t) \,.
  \label{eq:Twr}
\end{equation}
While this may be immediately inserted into Equation (\ref{eq:CPvar}) to obtain the variations in the closure phase, it is instructive to consider the limit in which there is a clear separation between the scales of $\bar{I}(\bmath{x})$ and the fluctuations.

For our purposes here, because the perturbations are expected to be small, the second term in Equation (\ref{eq:Twr}) may still be subdominant when this is realized in practice.  That is, we assume that $u\gg u_0$ where $u_0$ is the baseline length above which $\bar{V}(\u)$ begins to decrease substantially.  In the interest of concreteness we will assume that
\begin{equation}
  \bar{V}(\u) \approx
  V_0 
  \begin{cases}
    1 & u<u_0\\
    (u/u_0)^{-1} & u>u_0\,,
  \end{cases}
  \label{eq:Vbar}
\end{equation}
where the asymptotic power is typical of that arising from the power-law brightness decline in images for RIAFs. As a result,
\begin{equation}
  \left[\tilde{\bmath{\xi}}*\left(\frac{2\pi i\u}{\lambda} \bar{V}\right)\right](\u,t)
  \approx
  \frac{2\pi i u_0}{\lambda u} V_0\,
  \u\cdot\tilde{\bmath{\xi}}(\u,t)\,,
\label{eq:Vrefnoisemag}
\end{equation}
and $T(\u,t)\approx 1 + [2\pi i\u\cdot\bmath{\xi}(\u,t)/\lambda]$.  Inserting this into Equation (\ref{eq:CPvar}) yields
\begin{equation}
  \delta\Phi_{123}
  \approx
    \du\cdot\grad_{\bf u} \frac{2\pi \u\cdot\bmath{\xi}(\u,t)}{\lambda}\,.
\end{equation}
Between the characteristic scales of the quiescent image and the scattering screen, $\bmath{\xi}$ is roughly constant, giving
\begin{equation}
  \delta\Phi_{123} \approx
  \left(\frac{2 \pi u}{\lambda} \xi\right) \left(\frac{\delta u}{u}\right)\,.
\end{equation}
For $\delta u/u \approx 0.2$, appropriate for the CARMA-SMT vs. Hawaii baselines, reproducing the $\delta\Phi_{123}\approx0.07$ requires typical refractive distortions of $\xi\approx0.05 \lambda/u \approx 3~\muas$ on scales of $60~\muas$.

The intra-day timescale is consistent with recent models of a ``nearby'' scattering screen \citep{2014ApJ...780L...2B} motivated by observations of the recently discovered magnetar \citep{2013ApJ...770L..24K} with a velocity of $\approx30\,\km\,\s^{-1}$, similar to the velocity dispersion of stars in the disk.  It is also consistent, however, with a ``distant'' scattering screen \citep[e.g.,][]{1998ApJ...505..715L} assuming velocities of $\approx100\,\km\,\s^{-1}$, comparable to those expected in the bulge.  Thus, both the magnitude and timescale of the closure phase variations is consistent with an origin in the scattering screen.

\subsection{Simulated Strong Refractive Scattering} \label{sec:strong_scatt}
In the strong-scattering regime, i.e., where the scattering induces angular rearrangements on scales comparable to the structures in the image, the preceding perturbative analysis is insufficient.  Here we briefly describe an attempt to simulate this scattering and numerically infer the typical variations in the observed closure phases.

As in the weak case, strong refractive scattering in the interstellar medium produces stochastic fluctuations in images and, hence, in the visibility magnitudes and closure phases \citep{1989MNRAS.238..963N,1989MNRAS.238..995G,0004-637X-805-2-180}.  Although the strength of these fluctuations depends on properties of the scattering, the most significant uncertainties arise from the unknown source image: an extended source quenches the fluctuations in a manner that depends on its size and structure.  As in the perturbative case both nearby and distant scattering screen models are consistent with the observed intra-day variability.

\begin{figure}
  \begin{center}
  \includegraphics[width=\columnwidth]{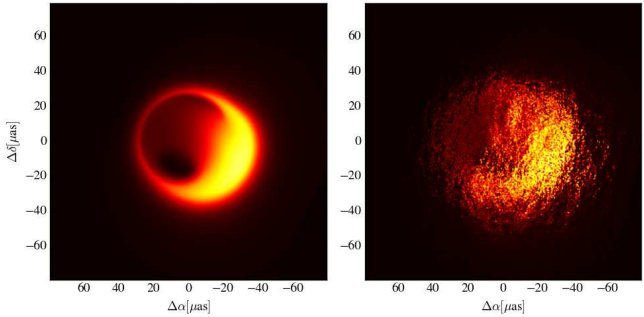}
  \caption{Input image (left) and example scattered image (right) for the best fit model.} \label{fig:scatt_ex}
  \end{center}
\end{figure}

\begin{figure}
  \begin{center}
    \includegraphics[width=\columnwidth]{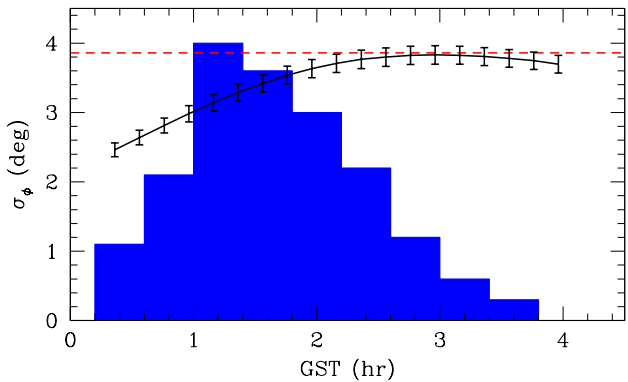}
  \end{center}
  \caption{Estimated root-mean square fluctuations in the closure phase on the SMT-CARMA-SMA baseline caused by refractive scattering as a function of observation time.  The $3.86^\circ$ standard deviation observed in the closure phases is shown by the red dashed line.  For reference, the distribution ($\times0.1$) of the employed closure phase measurements in time is shown in blue.}\label{fig:scatt_rms}
\end{figure}

Figure \ref{fig:scatt_ex} shows an example of the effects of refractive scattering on our best-fit model. The scattering kernel was taken from \citet{2006ApJ...648L.127B}, and we assumed an inner scale of $1.5\times10^4~\km$ for the turbulence. Although this inner scale is somewhat larger than expected for the interstellar medium, it simplifies the scattering simulation and has little effect (${\sim}10\%$) on the resulting refractive fluctuations. Following \citet{0004-637X-805-2-180}, we scattered the model image by first generating a random phase screen with $2^{13}{\times}2^{13}$ random phases and then shifting the unscattered image by the scaled gradient of the phase screen.

Figure \ref{fig:scatt_rms} shows the estimated root-mean-square (RMS) fluctuations of closure phase on the SMT-CARMA-SMA triangle, estimated by sampling the visibilities on an ensemble of scattered images.  These vary with time due to the time-variable orientation of the participating telescopes.  The times at which observations were made extend from 0.5~GST to 3.8~GST, with a median near 1.8~GST, suggesting a typical RMS near $3.5^\circ$.  This is very similar to the $3.86^\circ$ standard deviation observed implying that the bulk of the closure phase variation may be due to interstellar scattering.

\subsection{Accretion Flow Turbulence}
The impact of turbulence on the image is complicated by anisotropy and inhomogeneity as well as opacity within the emission region.  Here we will ignore these complications in the interest of obtaining a qualitative result, assuming an optically thick, homogeneous emission region, appropriate for the brightest component of the image of Sgr~A*.  In this case, the intensity is proportional to $n_e B^{-1/2}$, and hence fluctuations in the local electron density and magnetic field strength produce corresponding fluctuations in the intensity map.  For magnetohydrodynamic turbulence these are related via $\delta B/B = \alpha \delta n_e/n_e$ where $\alpha$ depends on the particular turbulence model under consideration (e.g., for fixed total pressure $\alpha=-1/2$, which we adopt), and thus $\delta I / I = (5/4) \delta n/n$.  Particle conservation relates the density variation to the fluid displacement field, i.e., assuming uniform density, $\delta n_e/n_e = - \grad\cdot\bmath{\xi}$.  While the turbulence is manifestly three dimensional, the observational consequences are dominated by the transverse mass redistribution, and hence we will concern ourselves only with the two-dimensional dimensional structure of the density (and magnetic field) on the sky.  Therefore, the resulting intensity map is given by
\begin{equation}
  I(\bmath{x},t)
  =
  \bar{I}(\bmath{x})
  \left[1 + \frac{5}{4} \grad\cdot\bmath{\xi}(\bmath{x},t)\right]\,.
\label{eq:Iaft}
\end{equation}
This is similar to the weakly refraction case, with the exception that now it is the divergence of $\bmath{\xi}$ that enters.

There is no compelling reason for a separation of scales between features in the quiescent image and the impact of turbulence.  Nevertheless, we will continue to make this assumption in the interest of computational expedience, providing only a qualitative assessment of the impact of turbulence on the closure phase variability.  As a result,
\begin{equation}
  T(\u,t) = 1 + \frac{5}{4 \bar{V}(\u)} \left[\left(\frac{2\pi i \u}{\lambda}\cdot\tilde{\bmath{\xi}}\right) *  \bar{V} \right](\u,t) \,.
  \label{eq:Taft}
\end{equation}
As before we will assume that $\bar{V}(\u)$ takes the approximate form in Equation (\ref{eq:Vbar}), in which case
\begin{equation}
\left[\left(\frac{2\pi i \u}{\lambda}\cdot\tilde{\bmath{\xi}}\right) *  \bar{V}\right](\u,t)
\approx
\frac{2\pi i}{\lambda} V_0 \, \u\cdot\bmath{\xi}(\u,t)\,.
\end{equation}
Inserting this into Equation (\ref{eq:CPvar}) then gives
\begin{equation}
  \begin{aligned}
  \delta\Phi_{123}
  &\approx
  \frac{5}{4} \du\cdot\grad_{\bf u} \frac{2\pi u \u\cdot\bmath{\xi}(\u,t)}{\lambda u_0}\\
  &\approx
  \frac{5}{2} \left(\frac{2\pi u}{\lambda}\xi\right) \left(\frac{u}{u_0}\right) \left(\frac{\delta u}{u}\right)\,.
  \end{aligned}
\end{equation}
Again, inserting typical scales: $\delta u/u\approx 0.2$, $u_0/u\approx0.2$, which requires $\xi\approx 0.003 \lambda/u$, i.e., turbulence displacements on scales of order $0.3\%$ of those probed by the Hawaii-SMT/CARMA baselines or $0.2~\muas$.  With the strong gravitational lensing, and therefore compression of the features within the disk around the photon ring, this implies turbulent displacements of order $5\%$ of the disk scale height, i.e., $5\%$ density variations.

The difference in physical scale from the weak refraction model is due entirely to the different coupling to the background intensity field combined with the assumed scale hierarchy.  For the weakly refracted case it is gradients of the quiescent intensity map that enter, which are limited to angular scales of $\lambda/u_0$;  for the accretion turbulence case it is gradients of the displacement field, which exist on scales of $\lambda/u$ and are thus better matched to the interferometric measurements.

The small turbulence amplitude required appears problematic initially.  However, this is the projected turbulence amplitude, obtained after integrating over the column, which will generically average down the impact of the turbulent variations.  Strong turbulence (i.e., 100\% local density variations) requires roughly $400$ contributing turbulent eddies to produce the required 5\% variations.  Since the photosphere volume is roughly $\pi r^3$,  where $r$ is the orbital radius, this implies only $\approx 5$ turbulent eddies per scale height (commensurate with $r$).

The timescale for strong accretion-driven turbulence driven variability is ostensibly on the order of the orbital periods, which is roughly $0.5~\hr$ at the ISCO of a non-spinning black hole in Sgr~A*.  However, this is a reasonably strong function of the radius, scaling as $r^{3/2}$.  This size is well matched to the projected scales being probed by the Hawaii-SMT baseline, which has a nominal resolution of $60~\muas$, corresponding to a linear scale of $10GM/c^2$, at which the orbital timescale is $1~\hr$.  For even moderately sub-Keplerian orbits this can grow to the scales needed to produce the observed inter-epoch variations.

\section{Analytical Marginalization over Closure Phase Shifts} \label{app:CPmarg}
Motivated by the theoretical expectation of inter-epoch shifts in the closure phases driven by small-scale image structure that we have ignored in our modeling, we permit each closure phase epoch a limited shift in the over-all closure phase values.  As with the flux renormalization this is an additional nuisance parameter that we will ultimately seek to maximize or marginalize over.  Because of its simplicity, we can do this analytically; here we present a formalism similar to that in Appendix A of \citet{2014ApJ...784....7B} describing how we do this.

The epoch-specific closure phase data contribution to the likelihood given a set of model parameter $\bmath{p}$, and therefore model closure phases of $\hat{\Phi}_j(\bmath{p})+\phi_{E}$, is given by the normal expression:
\begin{equation}
  L(\phi_1,\phi_2,\dots;\bmath{p})
  =
  N \prod_E \exp\left[- \sum_j \frac{\left(\Phi_{j,E}-\phi_E-\hat{\Phi}_j\right)^2}{2\sigma_{j,E}^2}\right]\,,
\end{equation}
where $N$ is a fixed normalization.  Maximizing $L$ with respect to $\phi_E$ gives in the usual way the most likely offset,
\begin{equation}
\phi_E^M = \Sigma_E^2 \sum_j \frac{\Phi_{j,E}-\hat{\Phi}_j}{\sigma_{j,E}^2}
\quad\text{where}\quad
\frac{1}{\Sigma_E^2} \equiv \sum_j \frac{1}{\sigma_{j,E}^2}
\,,
\end{equation}
corresponding the the likelihood
\begin{equation}
  L^M
  =
  N \prod_E \exp \left[
    -\frac{{\phi_E^M}^2}{2\Sigma_E^2}
    -\sum_j \frac{\left(\Phi_{j,E}-\hat{\Phi}_j\right)^2}{2\sigma_{j,E}^2}
    \right]\,.
\end{equation}
Note that in terms of $L^M$ and $\phi_E^M$, the likelihood is particular simple,
\begin{equation}
L = L^M \prod_E \exp\left[-\frac{(\phi_E-\phi_E^M)^2}{2\Sigma_E^2}\right]
\end{equation}

The marginalized likelihood requires some information about the prior on $\phi_E$, $\Pi(\phi_E)$.  Motivated by the models presented in Appendix \ref{app:CPvar}, here we assume that this is Gaussian.  The width, $\sigma_\Phi$, is indicative of the amplitude of the refraction or turbulence responsible for the inter-epoch closure phase fluctuations, and as described in Section \ref{sec:CIO} we estimate this empirically.  Thus, the marginalized likelihood is given by
\begin{equation}
  \begin{aligned}
    \bar{L}
    &=
    L^M \prod_E \int d\phi_E \Pi(\phi_E)
    \exp \left[ - \frac{\left(\phi_E-\phi_E^M\right)^2}{2\Sigma_E^2}  \right]\\
    &=
    \frac{L^M}{\sqrt{2\pi}\sigma_\Phi} \prod_E \int d\phi_E     
    \exp \left[ - \frac{\phi_E^2}{2\sigma_\Phi^2} - \frac{\left(\phi_E-\phi_E^M\right)^2}{2\Sigma_E^2} \right]\\    
    &=
    L^M \prod_E
    \frac{\Sigma_E}{\sqrt{\sigma_\Phi^2+\Sigma_E^2}}
    \exp\left[-\frac{{\phi_E^M}^2}{2(\sigma_\Phi^2+\Sigma_E^2)}\right]
  \end{aligned}
\end{equation}
This may be used to generate marginalized probability distributions as described in \citet{2014ApJ...784....7B}.

Finally, unlike the visibility magnitudes, the value of $\phi_E$ marginalized over $\Pi L$,
\begin{equation}
\bar{\phi}_E = \frac{\sigma_\Phi^2}{\sigma_\Phi^2+\Sigma_E^2} \phi_E^M\,,
\end{equation}
is not simply $\phi_E^M$ as a result of the imposition of a non-trivial prior.  Nevertheless, note that as the prior becomes weak, i.e., $\sigma_\Phi$ becomes large, the two become similar.

\end{appendix}

\acknowledgments A.E.B.~receives financial support from the Perimeter Institute for Theoretical Physics and the Natural Sciences and Engineering Research Council of Canada through a Discovery Grant.  Research at Perimeter Institute is supported by the Government of Canada through Industry Canada and by the Province of Ontario through the Ministry of Research and Innovation.

\bibliographystyle{apj}
\bibliography{gcpe3}

\end{document}